\RequirePackage{fix-cm}
\documentclass[smallextended]{svjour3}
\smartqed  

\usepackage{cite}
\usepackage{amsmath,amssymb,amsfonts}
\usepackage{algorithmic}
\usepackage{graphicx}
\usepackage{textcomp}
\usepackage{color, colortbl}
\usepackage{ifthen}
\usepackage{booktabs}
\usepackage{scalefnt}
\usepackage{color, colortbl}
\usepackage[flushleft]{threeparttable}
\usepackage[square,comma,numbers,sort]{natbib}
\usepackage{balance} 
\usepackage{multirow}
\usepackage{pifont}
\usepackage{subcaption}
\usepackage{caption}
\usepackage{lscape}
\usepackage{nicefrac}
\usepackage{hyperref}
\usepackage{pgfplots}
\usepackage{booktabs}
\usepackage{url}
\usepackage{fancybox}
\usepackage{multirow}
\usepackage{float}
\usepackage{array,graphicx}
\usepackage{subcaption}
\usepackage{listings}
\usepackage{pifont}
\usepackage{tikz}
\usepackage{balance}
\usepackage{xcolor}
\usepackage[most]{tcolorbox}
\usepgfplotslibrary{statistics}
\pgfplotsset{compat=1.15}

\newcommand{\Actions} {\textsc{GitHub Actions}}

\definecolor{purplish}{HTML}{D8D0E3}
\definecolor{purplishlight}{HTML}{EBE7F1}
\definecolor{purplishdark}{HTML}{168591}

\newtcolorbox[auto counter,number within=section]{rqbox}[2]{
    label=#2, nameref=#1,
    title=\small{#1}, 
    enhanced,
    attach boxed title to top left={yshift=-6pt, xshift=8pt},
    boxed title style={size=small,boxsep=1pt},
    colframe=purplishdark,colback=white,colbacktitle=purplishdark,
    boxsep=2pt,left=2pt,right=2pt,top=6pt,bottom=2pt,middle=2pt
}

\newcommand*{\img}[1]{%
    \raisebox{-.3\baselineskip}{%
        \includegraphics[
        height=\baselineskip,
        width=\baselineskip,
        keepaspectratio,
        ]{#1}%
    }%
}

\newcommand{\todo}[1]{}
\renewcommand{\todo}[1]{{\color{red} TODO: {#1}}}

\definecolor{Gray}{gray}{0.9}
\def\BibTeX{{\rm B\kern-.05em{\sc i\kern-.025em b}\kern-.08em
  T\kern-.1667em\lower.7ex\hbox{E}\kern-.125emX}}

\newcommand{\MyPara}[1]{\vspace{.2em}\noindent\textit{\textbf{#1}}\hspace{.3em}}

\journalname{Empirical Software Engineering manuscript}
\begin{document}

\title{\Actions{}: The Impact on the Pull Request Process}

\titlerunning{\Actions{}: The Impact on the Pull Request Process}        
\author{Mairieli Wessel \and
        Joseph Vargovich \and
        Marco A. Gerosa \and
        Christoph Treude
}

\authorrunning{Mairieli Wessel \and
        Joseph Vargovich \and
        Marco A. Gerosa \and
        Christoph Treude} 

\institute{Mairieli Wessel \at
            Radboud University, The Netherlands \\
            \email{mairieli.wessel@ru.nl} 
          \and 
          Joseph Vargovich \at
          Northern Arizona University, USA \\
          \email{jrv233@nau.edu}
            \and 
            Marco A. Gerosa \at
            Northern Arizona University, USA \\
          \email{Marco.Gerosa@nau.edu}
          \and 
          Christoph Treude \at
          University of Melbourne, Australia \\
          \email{christoph.treude@unimelb.edu.au}
}

\date{Received: date / Accepted: date}

\maketitle

\begin{abstract}
Software projects frequently use automation tools to perform repetitive activities in the distributed software development process. Recently, GitHub introduced \Actions{}, a feature providing automated workflows for software projects. Understanding and anticipating the effects of adopting such technology is important for planning and management. Our research investigates how projects use \Actions{}, what the developers discuss about them, and how project activity indicators change after their adoption. Our results indicate that 1,489 out of 5,000 most popular repositories (almost 30\% of our sample) adopt \Actions{} and that developers frequently ask for help implementing them. Our findings also suggest that the adoption of \Actions{} leads to more rejections of pull requests (PRs), more communication in accepted PRs and less communication in rejected PRs, fewer commits in accepted PRs and more commits in rejected PRs, and more time to accept a PR. We found similar results when segmenting our results by categories of \Actions{}. We suggest practitioners consider these effects when adopting \Actions{} on their projects.
\end{abstract}

\section{Introduction}
\label{sec:introduction}

Social coding platforms, such as GitHub, have changed the collaborative nature of open-source software development by integrating mechanisms such as issue reporting and pull requests into distributed version control tools~\cite{Dabbish2012,gousios2014exploratory}. This pull-based development workflow offers new opportunities for community engagement but increases the workload for repository maintainers, who need to communicate, review code, deal with contributor license agreement issues, explain project guidelines, run tests, and merge pull requests~\citep{Gousios2016}.

To reduce this intensive workload, developers often rely on automation tools to perform repetitive tasks, such as to check whether the code builds, the tests pass, and the contribution conforms to a defined style guide~\citep{kavaler2019tool}. GitHub projects adopt, for example, tools to support Continuous Integration and Continuous Delivery or Deployment (CI/CD)~\cite{zhao2017impact,cassee2020silent} and for code review~\cite{kavaler2019tool,wessel2020effects}. In recent years, development bots have been widely adopted to automate predefined tasks around pull requests~\cite{Wessel2018}. By automating part of the workflow, developers expect to increase both productivity and quality~\cite{Vasilescu2015}.

To further support automation, GitHub recently introduced \Actions{}\footnote{https://github.com/features/actions} (the feature was made available to the public in November 2019). \Actions{} allow the automation of tasks based on various triggers (e.g., commits, pull requests, issues, comments, etc.) and can be easily shared between repositories, automating aspects of how developers build, test, and deploy software projects. 

However, little is known about the impact on the project activities when adopting \Actions{} and the challenges imposed on the project development process. In this paper, we aim to understand how software developers use \Actions{} to automate their workflows and how the dynamics of pull requests of GitHub projects change following the adoption of \Actions{}. 

To achieve our goal, we address the following research questions:

\newcommand{\rqtextone}{How do open-source software projects use \Actions{}?}
\newcommand{\rqone}[2][]{
    \begin{rqbox}{\textbf{Research Question 1}}{#2}
        \rqtextone
        #1
    \end{rqbox}
}

\rqone{rq1}

We aim to understand how commonly repositories use \Actions{} and for what purposes. As a result of this analysis, we found a considerable number of active repositories (1,489 out of 5,000 repositories) adopted \Actions{}. This is a dramatic change when compared to the early adoption of \Actions{}~\cite{kinsman2021} (only 0.7\% of the studies repositories adopted it). Actions are spread across 20 categories, including utilities, continuous integration, code quality, and deployment.

\newcommand{\rqtexttwo}{How is the use of \Actions{} discussed by developers?}
\newcommand{\rqtwo}[2][]{
    \begin{rqbox}{\textbf{Research Question 2}}{#2}
        \rqtexttwo
        #1
    \end{rqbox}
}

\rqtwo{rq2}

To gain insight into how developers perceive \Actions{}, we manually analyzed a set of discussion threads and developer conversations on Discord that mention \Actions{}. We found distinct categories of discussions related to Actions, including help requests, the potential of using them, issues reproducing output with Actions, and plans to use \Actions{}.

\newcommand{\rqtextthree}{What is the impact of \Actions{} on the dynamics of pull requests?}
\newcommand{\rqthree}[2][]{
    \begin{rqbox}{\textbf{Research Question 3}}{#2}
        \rqtextthree
        #1
    \end{rqbox}
}

\rqthree{rq3}

In this research question, we investigate whether project activity indicators, such as the number of pull requests, comments, commits, and time to close pull requests change after \Actions{} adoption. We used a \textit{Regression Discontinuity Design} (RDD)~\cite{thistlethwaite1960regression} to model the effect of Action adoption across 662 projects that had adopted \Actions{} for at least 12 months. Our findings also suggest that the activity indicators change in opposite directions for accepted and rejected pull requests (PRs). Fewer pull requests are being accepted after adopting \Actions{}, and these pull requests usually have more comments and fewer commits. In contrast, there are more rejected pull requests, with fewer comments and more commits.

\newcommand{\rqtextfour}{How does the impact of \Actions{} differ across categories?}
\newcommand{\rqfour}[2][]{
    \begin{rqbox}{\textbf{Research Question 4}}{#2}
        \rqtextfour
        #1
    \end{rqbox}
}

\rqfour{rq4}

As Actions are diverse and might perform a diverse range of tasks on GitHub repositories, we also investigated whether the impact of \Actions{} differs across Action categories. The literature recommends employing a segmented analysis to further explain the general findings from statistical models~\cite{wessel2021quality}. In this research question, as in RQ3, we used a \textit{Regression Discontinuity Design} model to measure the impact of adoption in project indicators across the four most popular Action categories: Utilities, Continuous Integration, Code Quality, and Deployment. Results obtained in the segmented analysis were similar to the overall results (from RQ3), except for code quality Actions, which led to fewer rejected pull requests.

The main contributions of this paper are:

\begin{enumerate}
    \item Characterization of the usage of \Actions{}.
    \item An understanding of how developers discuss \Actions{}.
    \item An understanding of how \Actions{}' adoption impacts project activities.
\end{enumerate}

This paper extends our prior work~\cite{kinsman2021}, published at MSR 2021 (The Mining Software Repositories Conference), along two major dimensions: the data used and the analyses performed. The data used in this paper broadens our previous work in three major dimensions: time (24 vs. 12 months after \Actions{} were introduced), the number of unique Actions (973 vs. 708), and the dataset of projects used (5,000 most popular GitHub projects vs. RepoReapers dataset). In this extension, we also added RQ4 and included new regression discontinuity design analyses split by Action categories. 

\section{Workflow Automation with \Actions{}}

\begin{figure*}[!htbp]
\scriptsize
\centering
\includegraphics[scale=0.5]{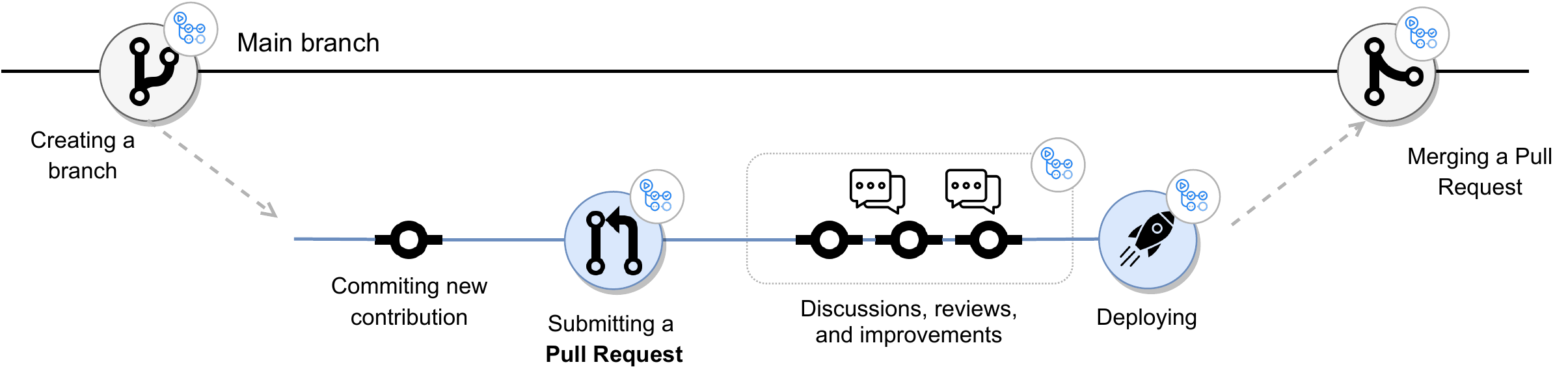}
\caption{GitHub workflow automation with \Actions{} (adapted from GitHub).}
\label{fig:flow}
\end{figure*}

\Actions{} is an event-driven API the GitHub platform provides to automate development workflows. \Actions{} can run a series of commands after a specified event has occurred. An event is a specific activity that triggers a workflow run, as shown in Figure~\ref{fig:flow} (see the \img{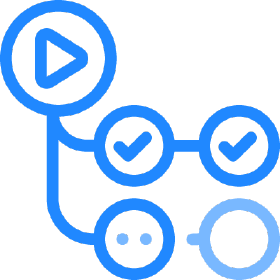} icon). For example, a workflow is triggered when a pull request is created for a repository or when a pull request is merged into the main branch. Workflows are defined in the \textbf{.github/workflows/} directory and use YAML syntax, having either a .yml or .yaml file extension. 

A workflow can contain one or more Actions. GitHub allows developers to build reusable components, called Actions. Developers create Docker and JavaScript Actions, and both require a metadata file to define the Action's inputs, outputs, and entry point.

After the successful execution of a workflow, the outputs can be displayed in different ways, such as through a \textit{GitHub Action bot}. Like many other bots on GitHub, this bot is implemented as a GitHub user that can submit code contributions, interact through comments, and merge or close pull requests~\cite{wessel2020inconvenient}.

\begin{figure}
     \centering
     \begin{subfigure}[b]{0.4\textwidth}
        \scriptsize
        \centering
        \includegraphics[width=1.5\textwidth]{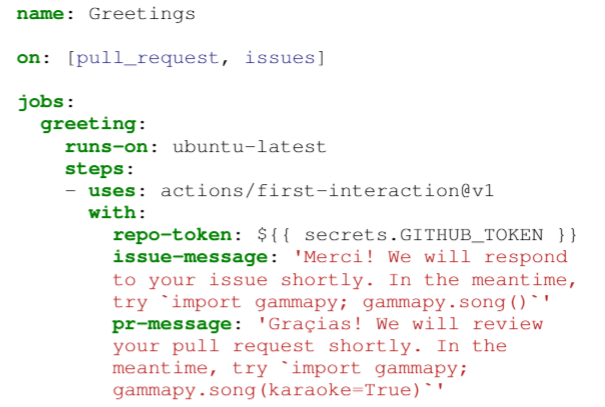}
        \caption{Greetings workflow of Gammapy -- \textit{greetings.yml}}
        \label{listing:greetings}
     \end{subfigure}
    \\
    \begin{subfigure}[b]{0.4\textwidth}
    \scriptsize
    \centering
    \includegraphics[width=1.5\textwidth]{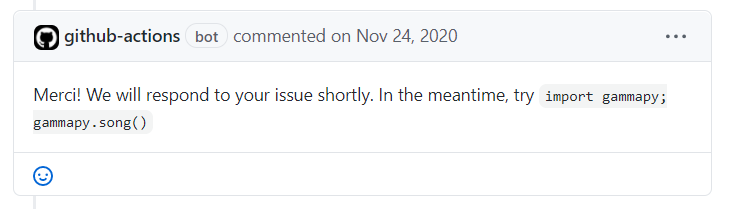}
    \caption{Example of \textit{github-actions} bot greeting a newcomer.}
    \label{fig:greetings-example}
    \end{subfigure}
\end{figure}

As an example of \Actions{} adoption, consider the case of the project \textit{Gammapy}\footnote{https://github.com/gammapy/gammapy}, an open-source Python package for gamma-ray astronomy. As of the 13$^{th}$ of November 2019, the \textit{Gammapy} community adopted a GitHub Action called \textit{First Interaction}\footnote{https://github.com/marketplace/actions/first-interaction}, which is responsible for identifying and welcoming newcomers when they create their first issue or open their first pull request on a project. As shown in Figure~\ref{listing:greetings}, \textit{Gammapy} created a workflow called \textit{Greeting} that both new pull requests and issues might trigger, as defined by the \textbf{on} keyword. The output of the \textit{First Interaction} Action is displayed through an issue/pull request comment posted by \textit{GitHub Action Bot} when a new contributor authors a new pull request or issue. An example of this Action interaction on a GitHub issue is shown in Figure~\ref{fig:greetings-example}.

Development bots and workflows that rely on \Actions{} are already used in hundreds of thousands of repositories, justifying the need for further studies on these automation mechanisms' evolution and impact on collaborative software development practices. Recently, developers published \Actions{} variants for many well-known bots (e.g., Coveralls, Codecov, Snyk), and these Actions are rapidly increasing in popularity~\cite{golzadeh2020groundtruth}.

\section{Related Work}

Previous work has investigated a variety of automation tools, including development bots, continuous integration/delivery, and \Actions{}.

\subsection{Development Bots}
Development bots have been proposed to automate technical and social aspects of software development activities~\cite{Lin2016}, such as communication and decision-making~\cite{Storey2016}. For example, on GitHub, bots are often integrated into the pull request workflow~\cite{Erlenhov2019} to perform a variety of tasks, including repairing bugs~\cite{Monperrus2019}, refactoring source code~\cite{Wyrich2019}, recommending tools to help developers~\cite{Brown2019}, and updating outdated dependencies~\cite{Mirhosseini2017}. Wessel et al.~\cite{Wessel2018} identified $13$ categories of development bots. Van Tonder and Le Goues~\cite{Tonder2019} believe development bots are a promising addition to a developer's toolkit as they bridge the gap between human software development and automated processes.

However, understanding the impact of development bots on human developers' interactions is a major challenge. Storey et al.~\cite{Storey2016} highlight that the way that development bots interact on pull requests can be disruptive and perceived as unwelcoming. Wessel et al.~\cite{wessel2021don} identified several challenges caused by bots in pull requests and theorized how human developers perceive annoying bot behaviors as noise on social coding platforms. Wessel et al.~\cite{wessel2020effects,wessel2021quality} also found that adopting code review bots changes team dynamics, for example, by increasing the number of monthly merged pull requests and decreasing communication among developers.

\subsection{Continuous Integration and Continuous Delivery}
Continuous Integration and Continuous Delivery (CI/CD) tools aim to bridge development and operation activities by automating the building, testing, and deployment of applications~\cite{duvall2007continuous}. These tools constantly compile incremental code changes made by developers, build software deliverables, run automated tests and verifications, and deploy applications to servers, improving software quality and productivity~\cite{duvall2007continuous}. Vasilescu et al.~\cite{Vasilescu2015} show that using CI leads to more pull requests being processed, and thus more pull requests being accepted or rejected. In the context of Computer Science education, Hu et al.~\cite{Hu2019} set up a continuous integration service on GitHub to provide feedback to students about code style and functionality. Prior work has also investigated the impact of CI and code review tools on GitHub projects~\cite{zhao2017impact,kavaler2019tool,cassee2020silent} across time. While Zhao et al.~\cite{zhao2017impact} and Cassee et al.~\cite{cassee2020silent} focused on the impact of the Travis CI tool's introduction in development practices, Kavaler et al.~\cite{kavaler2019tool} examined the impact of linters, dependency managers, and coverage reporter tools. A survey by Chen et al.~\cite{940726} reports that of the hundreds of billions of dollars spent on developer wages, up to 25\% accounts for fixing bugs~\cite{940726}. Continuous integration and other automation tools thus hold huge potential to further reduce human effort and costs by automatically fixing bugs.

\subsection{\Actions{}}
\Actions{} offer built-in support to automate parts of the software development workflows that exceed what CI/CD tools can achieve. 
Golzadeh et al.~\cite{golzadehrise} showed that, in 18 months of existence, \Actions{} had become the dominant CI service, covering more than half of all repositories with a CI. Software projects are still adjusting \Actions{} to their dynamics. Valenzuela-Toledo and Bergel~\cite{valenzuelaevolution} found 11 reasons for changing the \Actions{}' workflow. Saroar and Nayebi \cite{saroar23} conducted a survey to understand the motivations and best practices in using, developing, and debugging \Actions{}. Calefato et al. \cite{10.1145/3544902.3546636} identified a set of practices for using \Actions{} in projects related to machine learning-enabled systems. In a broader view, Decan et al. \cite{decan22actions} found that the reuse of actions is a common practice. Researchers are also starting to provide academic tools via \Actions{} to facilitate the integration with real projects. For example, Cordeiro et al.~\cite{cordeiro2021shaker} offer a GitHub Action for detecting flakiness in time-constrained tests. Finally, in a prior work~\cite{kinsman2021}, we investigate how developers use \Actions{} and how several activity indicators change after their adoption. We explain how this paper extends our prior work in Section~\ref{sec:introduction}. Chen et al.~\cite{chen2021let} also extended our prior work, finding that 22\% of popular projects adopt \Actions{} and that adoption correlates with project popularity and number of contributors and varies per programming language. They also found that after adopting \Actions{}, the number of commits, number of pull requests, issue latency, and pull request latency tend to decrease, while the number of issues closed tends to increase.

\section{Research Design}

This study aims to understand \Actions{} usage and the effects on GitHub projects. To achieve our goal, we employed a mixed-methods approach combining a time series analysis on a sample of open-source repositories and qualitative analysis of developers impressions about \Actions{}. We present our study design, data collection, and analysis procedures in the following.

\subsection{Selecting Projects}

We assembled a dataset of GitHub open-source projects that adopted \Actions{} at some point in their history. We started by selecting the 5,000 most-starred GitHub repositories. We used stars as a proxy for popularity. We then filtered this dataset to keep open-source software projects that had adopted at least one Action during their lifetime. To identify these projects, we retrieved data from the GitHub API using a Ruby toolkit called Octokit.rb.\footnote{\url{http://octokit.github.io/octokit.rb}}  To determine if the project used any Actions, we verified whether the repositories contained files in yaml format in the \textit{./github/workflows} directory. This filtered dataset comprised 1,489 projects.

\subsection{Analyzing the use of \Actions{}}

First, we collected and quantitatively analyzed the number of projects using \Actions{} and the number of \Actions{} per project (\textbf{RQ1}). We also automatically analyzed the workflow files of the studied projects, searching for the category, description, and whether GitHub verified the Action. We determined the Actions used within a workflow by extracting the `uses: ACTION@VERSION.' For example, in `uses: actions/first-interaction@v1' the \textit{First interaction}\footnote{\url{https://github.com/marketplace/actions/first-interaction}} was identified and extracted. In the case of multiple Actions in a single workflow, all of them were identified.

\subsection{Categorizing \Actions{} Discussions}

To answer \textbf{RQ2}, we manually investigated how \Actions{} were discussed in project-specific channels, including GitHub Discussions~\cite{hata2022github} and Discord chats~\cite{subash2022disco}.

\textbf{Filtering GitHub Discussions and Discord chats}. We started by investigating the GitHub Discussions on our selected projects. Out of the 5,000 repositories in our dataset, 897 (18\%) had the Discussions feature enabled at the time of data collection, and 830 (17\%) contained at least one Discussion thread. These 830 repositories account for 88,443 Discussion threads (minimum: 1, median: 22, maximum: 10,129), containing 326,033 posts. 
To complement our analysis, we have also considered developers' conversations on Discord, as they may use other communication channels to discuss \Actions{}. For this analysis, we used the DISCO dataset~\cite{subash2022disco}. This dataset consists of one-year public conversations on Discord of five software development communities (Python, Go, Clojure, and Racket).

Aiming for high precision rather than recall, we applied a strict filter to these GitHub Discussion posts and chat excerpts and selected only those with the exact string ``GitHub Action'' (case insensitive). We avoided searching for strings like ``.github/workflows/'' and ``workflow'', which tend to generate many false positives. An exploratory analysis of the DISCO dataset showed that strings like ``.github/workflows/'' are rarely mentioned, and ``workflow'' mostly appears in unrelated contexts.

This filtering step resulted in (i) 573 posts originating from 458 threads in 148 different repositories and (ii) 40 excerpts from two distinct communities (34 and 6 excerpts from Python and Go, respectively).

\textbf{Qualitative analysis.} We applied qualitative coding to the 458 threads to understand how developers discuss \Actions{}. One author developed a preliminary coding schema based on a random sample of 20 threads, which was refined through discussions with all authors. Two authors then independently coded another set of 20 threads and measured inter-rater agreement. Based on achieving an `almost perfect' agreement (Cohen's $\kappa = 0.939$~\cite{mchugh2012interrater}) and resolving disagreements through discussion, the same two authors divided the remaining threads equally among them and completed the annotation of all 458 threads. We also applied qualitative coding to the 40 chat excerpts from the DISCO dataset. Two authors then independently coded all chat excerpts based on the defined code schema and measured inter-rater agreement (Cohen's $\kappa = 1$). Section~\ref{sec:rq2-answer} reports the coding schema and the detailed results for both Discussion threads and Discord conversations.

\subsection{Time series analysis}

We conducted a time series analysis to answer \textbf{RQ3} and \textbf{RQ4}. We collected longitudinal data for different outcome variables and treated the adoption of \Actions{} by each project in our dataset as an ``intervention''. This way, we could align all the time series of project-level outcome variables on the intervention date and compare their trends before and after adopting \Actions{}. The following subsections detail the steps involved, from aggregating the project variables to running the statistical models.

\subsubsection{Aggregating project variables}

We gathered Action data within an observation period of 12 months before and 12 months after the Action adoption within each project. Similar to previous work~\cite{zhao2017impact,wessel2020effects,cassee2020silent,kinsman2021}, we exclude 30 days around the Action adoption date to avoid the influence of the instability caused during this period. Afterward, we aggregated individual pull request data into monthly periods, considering 12 months before and after the Action introduction. Next, we checked the activity level of the candidate projects, since many projects on GitHub are inactive~\cite{gousios2014exploratory}. Our data set comprises 662 active projects that had been using at least one GitHub Action for 12 months.

We focused on the same pull request-related variables as in previous work \cite{wessel2020effects,kinsman2021}:

\MyPara{Merged/non-merged pull requests:} the number of monthly contributions (pull requests) that have been merged (accepted) or closed but not merged (rejected) into the project, computed over all closed pull requests in each time frame.

\MyPara{Comments on merged/non-merged pull requests:} the median number of monthly comments computed over all merged and non-merged pull requests in each time frame.

\MyPara{Commits of merged/non-merged pull requests:} the median of monthly commits computed over all merged and non-merged pull requests in each time frame.

\MyPara{Time to merge/time to close pull requests:} the median of monthly pull request latency (in hours), computed as the difference between the time when the pull request was closed and the time when it was opened. The median is computed using all merged and non-merged pull requests in each time frame.

Based on previous work~\cite{cassee2020silent,zhao2017impact,wessel2020effects,kinsman2021}, we also collected six known covariates for each project:

\MyPara{Project name:} the name of the project to which the pull request belongs. This name is used to uniquely identify the project on GitHub.

\MyPara{Programming language:} the primary project programming language, as automatically provided by GitHub.

\MyPara{Time since the first pull request:} in months, computed since the earliest recorded pull request in the project's history. We use this variable to capture the project's maturity regarding its use of pull requests.

\MyPara{Total number of pull request authors:} we count how many contributors submitted pull requests to the project as a proxy for the community size of a project.

\MyPara{Total number of commits:} we compute the total number of commits as a proxy for the activity level of a project.

\MyPara{Number of pull requests opened:} the number of monthly contributions (pull requests) received in each time frame. We expect that projects with a high number of contributions also observe a high number of comments, latency, commits, and merged and non-merged contributions.

\subsubsection{Statistical Approach}
\label{sec-statistical-modeling}

We modeled the effect of GitHub Action adoption over time across GitHub repositories using a Regression Discontinuity Design (RDD)~\cite{thistlethwaite1960regression,imbens2008regression}, following the work of Wessel et al. \cite{wessel2020effects}. RDD is a technique used to model the extent of a discontinuity at the moment of intervention and long after the intervention. The technique assumes that if the intervention does not affect the outcome, there would be no discontinuity, and the outcome would be continuous over time~\cite{quasiexperimentation}. 
The statistical model behind RDD is
\begin{equation*}
\begin{split}
y_{i} =&\: \alpha + \beta\cdot \mbox{\textit{time}}_{i} + \gamma\cdot \mbox{\textit{intervention}}_{i} \: + \\& \delta\cdot \mbox{\textit{time\_after\_intervention}}_{i} \: + \eta\cdot \mbox{controls}_{i} + \varepsilon_{i}
\end{split}
\end{equation*}
where $i$ indicates the observations for a given project.

To model the passage of time as well as the GitHub Action introduction, we rely on three variables: \textit{time}, \textit{time after intervention}, and \textit{intervention}. The \textit{time} variable is measured as months at the time $j$ from the start to the end of our observation period for each project.

The \textit{intervention} variable is a binary value used to indicate whether the time $j$ occurs before ($\mbox{\textit{intervention}}=0$) or after the ($\mbox{\textit{intervention}}=1$) adoption event. The \textit{time\_after\_intervention} variable counts the number of months at time $j$ since the Action adoption, and the variable is set to 0 before adoption. The $\mbox{\textit{controls}}_{i}$ variables enable the analysis of Action adoption effects rather than confounding the effects that influence the dependent variables. For observations before the intervention, holding controls constant, the resulting regression line has a slope of $\beta$, and after the intervention $\beta+\delta$. The size of the intervention effect is measured as the difference equal to $\gamma$ between the two regression values of $y_{i}$ at the moment of the intervention. 

Considering that in \textbf{RQ3} we are interested in the effects of \Actions{} on the monthly trend of the number of pull requests, number of comments, number of commits, and time to close for both merged and non-merged pull requests, we fitted eight models (4 variables $\times$ 2 cases). In \textbf{RQ4}, we measured the impact of adoption for the same variables across the four most popular Action categories in our filtered dataset: utilities, continuous integration, code quality, and deployment. We selected projects that have adopted one or more of the four categories. In cases where a project employs multiple Actions, the project is considered in the analysis of multiple Action categories. Therefore, we fitted thirty-two models (4 variables $\times$ 2 cases $\times$ 4 categories).

To balance false positives and false negatives, we report the corrected p-values after applying multiple corrections using the method of Benjamini and Hochberg~\cite{benjamini1995controlling}. We implemented the RDD models as a mixed-effects linear regression using the R package \textit{lmerTest}~\cite{kuznetsova2017lmertest}. We modeled \textit{project name} and \textit{programming language} as random effects~\cite{galecki2013linear} to capture project-to-project and language-to-language variability~\cite{zhao2017impact}. We evaluate the model fit using \textit{marginal} $(R^2_m)$ and \textit{conditional} $(R^2_c)$ scores, as described by Nakagawa and Schielzeth~\cite{nakagawa2013general}. The $R^2_m$ can be interpreted as the variance explained by the fixed effects alone, and $R^2_c$ as the variance explained by the fixed and random effects together.

In mixed-effects regression, the variables used to model the intervention and the other fixed effects are aggregated across all projects, resulting in coefficients useful for interpretation. The interpretation of these regression coefficients supports the discussion of the intervention and its effects, if any. Thus, we report the significant coefficients ($p < 0.05$) in the regression and their variance, obtained using ANOVA. In addition, we \textit{log} transform the fixed effects and dependent variables that have high variance~\cite{sheather2009modern}. We also account for multicollinearity, excluding any fixed effects for which the variance inflation factor (VIF) is higher than $5$~\cite{sheather2009modern}.

\section{Results}

In the following, we report the results of our study per research question.

\subsection{How do OSS projects use \Actions{}? (RQ1)}
\label{howmany}

\begin{figure}
\centering
\begin{tikzpicture}
\begin{axis}[y=1cm, try min ticks=2, ytick=5, xmode=log, xtick={1,5,10,50,100}, xticklabels={1,5,10,50,100}]
\addplot[black, mark=o, boxplot, color=black]
table[row sep=\\,y index=0] {
data\\
2\\  12\\  2\\  3\\  3\\  3\\  7\\  6\\  2\\  7\\  4\\  4\\  4\\  13\\  2\\  1\\  2\\  7\\  2\\  2\\  6\\  2\\  3\\  3\\  12\\  4\\  2\\  4\\  3\\  3\\  2\\  6\\  3\\  6\\  2\\  1\\  6\\  4\\  3\\  2\\  7\\  4\\  5\\  9\\  2\\  2\\  2\\  3\\  7\\  1\\  2\\  5\\  4\\  6\\  2\\  1\\  1\\  8\\  1\\  1\\  4\\  1\\  2\\  4\\  2\\  2\\  2\\  3\\  6\\  2\\  19\\  12\\  3\\  4\\  16\\  10\\  23\\  14\\  9\\  6\\  4\\  5\\  2\\  5\\  8\\  4\\  3\\  16\\  9\\  2\\  9\\  14\\  18\\  3\\  4\\  4\\  4\\  2\\  1\\  2\\  8\\  7\\  4\\  1\\  4\\  2\\  4\\  2\\  3\\  2\\  8\\  2\\  7\\  4\\  6\\  2\\  3\\  2\\  4\\  10\\  10\\  3\\  2\\  2\\  12\\  4\\  4\\  2\\  4\\  3\\  1\\  9\\  2\\  15\\  3\\  3\\  6\\  1\\  7\\  3\\  2\\  4\\  2\\  4\\  3\\  4\\  7\\  1\\  1\\  8\\  5\\  10\\  7\\  1\\  3\\  2\\  4\\  2\\  26\\  2\\  2\\  3\\  8\\  2\\  2\\  3\\  9\\  12\\  1\\  5\\  2\\  2\\  2\\  4\\  3\\  7\\  2\\  6\\  2\\  3\\  6\\  3\\  3\\  7\\  1\\  3\\  4\\  2\\  8\\  8\\  7\\  2\\  2\\  4\\  3\\  3\\  1\\  7\\  7\\  5\\  1\\  2\\  2\\  1\\  8\\  12\\  12\\  5\\  1\\  5\\  4\\  11\\  1\\  9\\  7\\  3\\  21\\  8\\  7\\  5\\  3\\  7\\  5\\  10\\  4\\  1\\  7\\  3\\  2\\  4\\  6\\  9\\  8\\  9\\  12\\  6\\  3\\  2\\  8\\  2\\  7\\  4\\  7\\  4\\  8\\  7\\  1\\  4\\  2\\  3\\  4\\  6\\  3\\  2\\  1\\  2\\  1\\  3\\  6\\  2\\  2\\  1\\  6\\  2\\  5\\  3\\  6\\  1\\  2\\  2\\  3\\  3\\  3\\  2\\  10\\  3\\  7\\  3\\  2\\  2\\  6\\  3\\  1\\  2\\  3\\  15\\  2\\  6\\  6\\  2\\  8\\  7\\  9\\  2\\  1\\  2\\  8\\  7\\  8\\  8\\  8\\  8\\  8\\  8\\  8\\  8\\  8\\  8\\  6\\  13\\  4\\  8\\  6\\  9\\  6\\  7\\  2\\  11\\  4\\  1\\  5\\  2\\  4\\  2\\  2\\  4\\  4\\  14\\  9\\  4\\  11\\  2\\  5\\  8\\  5\\  2\\  2\\  11\\  5\\  10\\  12\\  4\\  6\\  2\\  14\\  3\\  5\\  5\\  4\\  2\\  4\\  2\\  6\\  2\\  3\\  2\\  10\\  6\\  2\\  2\\  2\\  3\\  2\\  2\\  10\\  7\\  6\\  3\\  4\\  10\\  2\\  2\\  7\\  1\\  6\\  4\\  12\\  2\\  3\\  1\\  3\\  10\\  2\\  1\\  22\\  2\\  4\\  4\\  3\\  6\\  5\\  7\\  3\\  3\\  5\\  4\\  6\\  4\\  7\\  3\\  5\\  4\\  6\\  7\\  2\\  1\\  7\\  7\\  2\\  2\\  2\\  9\\  5\\  2\\  8\\  2\\  5\\  5\\  6\\  6\\  5\\  2\\  3\\  6\\  6\\  3\\  3\\  13\\  2\\  3\\  2\\  3\\  2\\  3\\  7\\  4\\  3\\  4\\  1\\  3\\  2\\  4\\  3\\  5\\  2\\  11\\  1\\  2\\  8\\  4\\  1\\  16\\  4\\  3\\  5\\  3\\  2\\  3\\  6\\  2\\  2\\  2\\  2\\  4\\  2\\  2\\  2\\  2\\  2\\  7\\  4\\  8\\  23\\  15\\  6\\  3\\  1\\  5\\  2\\  9\\  10\\  4\\  2\\  46\\  3\\  3\\  5\\  5\\  4\\  4\\  4\\  3\\  17\\  8\\  5\\  3\\  4\\  3\\  7\\  13\\  2\\  12\\  5\\  2\\  2\\  3\\  5\\  4\\  2\\  5\\  5\\  3\\  3\\  7\\  2\\  3\\  7\\  1\\  4\\  2\\  7\\  2\\  2\\  6\\  2\\  6\\  4\\  2\\  3\\  4\\  4\\  5\\  5\\  5\\  5\\  2\\  1\\  2\\  8\\  4\\  3\\  10\\  3\\  2\\  3\\  2\\  6\\  5\\  4\\  4\\  11\\  7\\  9\\  1\\  6\\  4\\  6\\  5\\  4\\  2\\  6\\  2\\  6\\  3\\  3\\  4\\  5\\  5\\  2\\  3\\  7\\  2\\  3\\  2\\  3\\  3\\  6\\  2\\  16\\  2\\  7\\  8\\  2\\  2\\  3\\  3\\  10\\  9\\  30\\  4\\  4\\  4\\  5\\  3\\  3\\  3\\  2\\  2\\  7\\  6\\  4\\  4\\  4\\  3\\  8\\  6\\  18\\  11\\  7\\  19\\  22\\  10\\  7\\  8\\  2\\  6\\  1\\  5\\  5\\  3\\  5\\  2\\  8\\  6\\  3\\  2\\  4\\  2\\  4\\  10\\  1\\  4\\  10\\  3\\  1\\  5\\  2\\  2\\  2\\  5\\  7\\  2\\  5\\  2\\  3\\  7\\  3\\  2\\  2\\  7\\  7\\  7\\  2\\  3\\  4\\  6\\  3\\  5\\  3\\  5\\  8\\  4\\  2\\  2\\  1\\  8\\  5\\  4\\  5\\  14\\  3\\  17\\  7\\  2\\  4\\  4\\  6\\  5\\  7\\  11\\  1\\  1\\  1\\  2\\  2\\  3\\  7\\  3\\  2\\  4\\  2\\  8\\  4\\  3\\  8\\  7\\  2\\  11\\  2\\  8\\  6\\  14\\  3\\  3\\  5\\  4\\  2\\  10\\  1\\  6\\  9\\  7\\  2\\  2\\  1\\  3\\  7\\  5\\  4\\  3\\  7\\  8\\  6\\  2\\  2\\  5\\  2\\  3\\  3\\  1\\  3\\  7\\  15\\  10\\  4\\  2\\  7\\  3\\  1\\  8\\  2\\  4\\  5\\  6\\  3\\  3\\  3\\  3\\  5\\  4\\  6\\  5\\  5\\  14\\  7\\  6\\  4\\  3\\  1\\  2\\  1\\  4\\  7\\  9\\  3\\  9\\  2\\  2\\  4\\  3\\  6\\  1\\  6\\  2\\  2\\  5\\  2\\  7\\  3\\  4\\  3\\  6\\  1\\  4\\  4\\  12\\  3\\  3\\  6\\  7\\  9\\  4\\  5\\  7\\  12\\  2\\  5\\  1\\  3\\  2\\  3\\  1\\  1\\  2\\  2\\  2\\  6\\  4\\  16\\  4\\  5\\  3\\  6\\  2\\  1\\  11\\  1\\  1\\  4\\  2\\  7\\  9\\  4\\  6\\  5\\  1\\  8\\  3\\  13\\  3\\  1\\  3\\  2\\  2\\  5\\  4\\  8\\  2\\  1\\  8\\  8\\  9\\  4\\  2\\  3\\  17\\  2\\  2\\  1\\  10\\  1\\  6\\  2\\  5\\  4\\  3\\  13\\  9\\  3\\  2\\  8\\  2\\  1\\  4\\  5\\  8\\  11\\  6\\  2\\  3\\  2\\  12\\  2\\  3\\  2\\  4\\  6\\  6\\  5\\  4\\  2\\  3\\  5\\  4\\  3\\  3\\  4\\  4\\  4\\  9\\  2\\  4\\  24\\  4\\  5\\  4\\  4\\  10\\  13\\  5\\  2\\  2\\  8\\  4\\  3\\  9\\  7\\  9\\  2\\  3\\  8\\  13\\  3\\  4\\  6\\  4\\  2\\  3\\  3\\  2\\  5\\  12\\  3\\  3\\  6\\  3\\  2\\  3\\  3\\  3\\  4\\  7\\  16\\  1\\  4\\  7\\  5\\  2\\  4\\  5\\  3\\  2\\  3\\  4\\  6\\  4\\  4\\  7\\  6\\  11\\  13\\  2\\  5\\  4\\  7\\  7\\  6\\  14\\  11\\  8\\  6\\  4\\  4\\  4\\  4\\  13\\  13\\  2\\  7\\  4\\  2\\  6\\  1\\  8\\  1\\  3\\  5\\  6\\  7\\  7\\  8\\  12\\  5\\  5\\  2\\  3\\  6\\  5\\  2\\  4\\  4\\  6\\  2\\  3\\  5\\  7\\  3\\  4\\  1\\  3\\  6\\  4\\  1\\  1\\  3\\  12\\  4\\  10\\  8\\  2\\  14\\  4\\  8\\  3\\  2\\  2\\  5\\  3\\  2\\  6\\  7\\  4\\  2\\  6\\  1\\  5\\  8\\  9\\  6\\  7\\  3\\  8\\  6\\  6\\  5\\  9\\  4\\  3\\  3\\  12\\  6\\  24\\  2\\  2\\  3\\  1\\  4\\  2\\  4\\  4\\  2\\  3\\  11\\  7\\  6\\  2\\  5\\  7\\  2\\  4\\  2\\  5\\  4\\  3\\  5\\  2\\  4\\  2\\  2\\  12\\  2\\  6\\  5\\  3\\  2\\  1\\  5\\  3\\  6\\  4\\  2\\  6\\  3\\  2\\  6\\  5\\  4\\  3\\  4\\  4\\  7\\  5\\  4\\  4\\  4\\  3\\  3\\  5\\  1\\  3\\  8\\  4\\  5\\  1\\  5\\  8\\  6\\  13\\  3\\  1\\  5\\  2\\  6\\  3\\  2\\  7\\  1\\  9\\  2\\  10\\  2\\  4\\  2\\  2\\  1\\  1\\  1\\  1\\  2\\  3\\  8\\  5\\  5\\  1\\  4\\  1\\  3\\  9\\  2\\  7\\  3\\  2\\  4\\  3\\  10\\  4\\  6\\  3\\  1\\  8\\  1\\  7\\  8\\  8\\  9\\  2\\  2\\  6\\  2\\  3\\  3\\  3\\  3\\  3\\  3\\  3\\  3\\  3\\  3\\  4\\  3\\  3\\  3\\  3\\  3\\  4\\  5\\  6\\  7\\  10\\  5\\  2\\  2\\  2\\  3\\  3\\  6\\  6\\  4\\  5\\  6\\  3\\  1\\  4\\  7\\  2\\  2\\  3\\  4\\  3\\  6\\  6\\  4\\  2\\  4\\  4\\  4\\  3\\  11\\  5\\  3\\  2\\  3\\  3\\  2\\  2\\  2\\  3\\  2\\  2\\  2\\  4\\  4\\  4\\  1\\  5\\  3\\  7\\  3\\  4\\  2\\  1\\  2\\  2\\  2\\  4\\  1\\  2\\  1\\  2\\  1\\  21\\  3\\  2\\  5\\  4\\  5\\  4\\  3\\  6\\  7\\  8\\  2\\  2\\  1\\  2\\  9\\  15\\  13\\  2\\  2\\  5\\  4\\  5\\  8\\  4\\  8\\  4\\  2\\  6\\  3\\  3\\  4\\  6\\  5\\  2\\  2\\  7\\  6\\  6\\  3\\  7\\  2\\  2\\  5\\  4\\  8\\  4\\  2\\  1\\  8\\  3\\  18\\  10\\  7\\  4\\  3\\  5\\  7\\  11\\  4\\  17\\  8\\  8\\  2\\  4\\  4\\  4\\  21\\  1\\  3\\  1\\  13\\  15\\  4\\  4\\  8\\  19\\  2\\  3\\  8\\  1\\  2\\  2\\  1\\  10\\  1\\  7\\  6\\  6\\  4\\  5\\  3\\  4\\  3\\  6\\  2\\  2\\  2\\  3\\  1\\  10\\  1\\  6\\  2\\  2\\  2\\  1\\  1\\  2\\  4\\  3\\  2\\  1\\  4\\  2\\  3\\  6\\  2\\  10\\  2\\  3\\  2\\  1\\  2\\  2\\  9\\  2\\  2\\  8\\  8\\  2\\  4\\  3\\  5\\  4\\  6\\  6\\  7\\  4\\  2\\  2\\  6\\  1\\  1\\  8\\  11\\  4\\  1\\  3\\  3\\  2\\  2\\  2\\  4\\  1\\  4\\  3\\  7\\  3\\  6\\  11\\  3\\  9\\  4\\  3\\  4\\  3\\  8\\  7\\  4\\  5\\  3\\  3\\  7\\  4\\  2\\  5\\  6\\  2\\  3\\  4\\  6\\  5\\  2\\  4\\  4\\  10\\  13\\  4\\  5\\  2\\  2\\  6\\  2\\  5\\  9\\  13\\  4\\  1\\  1\\  3\\  6\\  4\\  7\\  1\\  16\\  12\\  8\\  3\\  5\\  3\\  3\\  3\\  5\\  5\\  7\\  2\\  2\\  2\\  2\\  4\\  9\\  5\\  1\\  2\\  4\\  5\\  6\\  3\\  8\\  1\\  4\\  3\\  8\\  9\\  3\\  6\\  2\\  2\\  3\\  6\\  3\\  1\\  2\\  2\\  8\\  9\\  4\\  
};
\end{axis}
\end{tikzpicture}
\caption{Number of Actions per repository (log scale)}
\label{fig:actions-per-repo}
\end{figure}
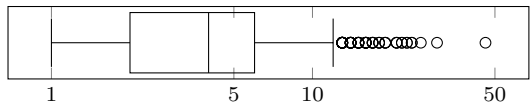

Analyzing the set of 5,000 repositories, we identified 1,489 (29.8\%) open-source software projects that had adopted at least one GitHub Action at the time of our data collection. As the box plot in Figure~\ref{fig:actions-per-repo} shows, many of these repositories adopt more than one Action, with a median value of four and a maximum of 46.

In these repositories, we found 973 distinct predefined \Actions{}. We collected data from each Action's repository and the GitHub Marketplace\footnote{https://github.com/marketplace?type=actions} page to categorize these \Actions{}. If published in the marketplace, an Action is classified into 1--2 categories by the publisher. Table~\ref{tab:categories} presents the categorization of \Actions{} we found. Note that the percentages do not add up to 100, since about half of the \Actions{} are assigned to two categories, a primary and a secondary.

\begin{table}[!htbp]
\centering
\caption[Categorization of \Actions{} found in our sample]{Categorization of \Actions{} found in our sample.}
\begin{tabular}{lrr}
\hline 
\textbf{\Actions{}' Categories}    & \textbf{\# of \Actions{}} & \textbf{\%}  \\ \hline
Utilities              & 243 & 24.97\% \\
Continuous integration & 241 & 24.77\% \\
Deployment             & 94  & 9.66\%  \\
Publishing             & 82  & 8.43\%  \\
Code quality           & 75  & 7.71\%  \\
Open Source management & 61  & 6.27\%  \\
Code review            & 58  & 5.96\%  \\
Testing                & 57  & 5.86\%  \\
Project management     & 49  & 5.04\%  \\
Dependency management  & 47  & 4.83\%  \\
Container CI           & 34  & 3.49\%  \\
Chat                   & 23  & 2.36\%  \\
Reporting              & 23  & 2.36\%  \\
Security               & 18  & 1.85\%  \\
Monitoring             & 9   & 0.92\%  \\
AI Assisted            & 7   & 0.72\%  \\
Code search            & 7   & 0.72\%  \\
Community              & 7   & 0.72\%  \\
Support                & 7   & 0.72\%  \\
Mobile CI              & 5   & 0.51\%  \\
API management         & 4   & 0.41\%  \\
Desktop tools          & 4   & 0.41\%  \\
Localization           & 2   & 0.21\%  \\
IDEs                   & 2   & 0.21\%  \\
Mobile                 & 2   & 0.21\%  \\
Code Scanning Ready    & 1   & 0.10\%  \\
Backup Utilities       & 1   & 0.10\%  \\ \hline
\textbf{total \Actions{}} & \textbf{973}     & \textbf{119.53} \\ \hline
\end{tabular}
\label{tab:categories}
\end{table}

The five most frequent categories of \Actions{} are:

\textbf{Utilities:} \Actions{} created to automate diverse steps of the development workflow on the GitHub platform, often in support of other \Actions{}. For example, the \textit{Read Properties} Action inspects Java \textit{.properties} files looking for predefined properties. Another example of a utility Action is \textit{Replace string}, which replaces strings that match predefined regular expressions.

\textbf{Continuous integration:} \Actions{} responsible for running the CI pipeline and notifying contributors of test failures in CI tools (e.g., Retry Step, Chef Delivery).

\textbf{Deployment:} \Actions{} designed to build and deploy the application upon request. One example is the Action called \textit{Jekyll Deploy}, responsible for building and deploying the Jekyll site to GitHub Pages.

\textbf{Publishing:} \Actions{} responsible for automatically publishing packages to the registry. For example, \textit{Action For Semantic Release} is an Action that leverages \textit{semantic-release} to fully automate the package release workflow, determining the next version number, generating the release notes, and publishing the package.

\textbf{Code quality:} \Actions{} that analyze source code (e.g., code style, code coverage, code quality, and smells) submitted through pull requests and give feedback to developers via GitHub checks or comments.

In addition, we found that 42 (5.93\%) out of 973 \Actions{} are verified by GitHub. Creators are verified if they have an existing relationship with GitHub, and GitHub works closely with the creator to generate these \Actions{}.

\begin{table}[!htbp]
\centering
\caption[Most-used \Actions{} across repositories.]{Most-used \Actions{} across repositories.}
\begin{tabular}{lrr}
\hline 
\textbf{Action}    & \textbf{\# of Repositories} & \textbf{\% out of 1,489}  \\ \hline

actions/checkout             & 1,442 & 96.84\% \\
actions/cache                & 485  & 32.57\% \\
actions/setup-node           & 461  & 30.96\% \\
actions/upload-artifact      & 353  & 23.71\% \\
actions/setup-python         & 279  & 18.74\% \\
github/codeql-action/init    & 156  & 10.48\% \\
github/codeql-action/analyze & 156  & 10.48\% \\
actions/setup-java           & 152  & 10.21\% \\
actions/download-artifact    & 148  & 9.94\%  \\
codecov/codecov-action       & 143  & 9.60\%  \\
\hline
\end{tabular}
\label{tab:actions}
\end{table}

Table~\ref{tab:actions} shows the ten most popular \Actions{}. The most popular one, actions/checkout is used by the vast majority (97\%) of repositories that have adopted at least one \Actions{}. The five most popular \Actions{} are the following:

\MyPara{actions/checkout:} A verified utility Action that checks out a repository under \$GITHUB\_WORKSPACE. Therefore, a workflow can access the repository for further workflow tasks.

\MyPara{actions/cache:} A verified utility and dependency management Action that allows caching dependencies and building outputs to improve workflow execution time.

\MyPara{actions/setup-node:} A verified utility Action that sets up a Node.js environment for use in a workflow, allowing users to specify a Node.js version.

\MyPara{actions/upload-artifact:} A verified utility Action that uploads artifacts from a workflow, allowing developers to share data between jobs and store data once a workflow is complete.

\MyPara{actions/setup-python:} A verified utility Action that sets up a Python environment for use in a workflow, allowing the use of Python features and commands.

\rqone[
    \tcblower
    Out of 5,000 GitHub repositories, 1,489 (29.8\%) adopted the \Actions{} feature, with a median of four \Actions{} used per repository. We found 973 unique predefined \Actions{} being used within the workflows. These \Actions{} are spread across 27 categories. The most recurrent ones are utilities, continuous integration, and deployment.
    \newline \newline
    \textbf{Comparison to our previous work}: In our previous work, we found that only 0.7\% of repositories considered in our analysis had adopted \Actions{}. This number has changed dramatically, with \Actions{} now having found much more widespread adoption.
]{}

\subsection{How is the use of \Actions{} discussed by developers? (RQ2)}
\label{sec:rq2-answer}

We categorized 458 GitHub Discussion threads and 40 developers' conversation excerpts containing the phrase ``GitHub Action''. Table~\ref{tab:issues} shows an overview of this categorization, indicating how many threads and excerpts we found in each category. We present the categories in the following.

\begin{table}[!htbp]
\centering
\caption[Categorization of discussions]{Categorization of Discussion threads and developers conversations on Discord.}
\begin{tabular}{p{5cm}@{}rrr}
\hline
\textbf{Category} && \textbf{\# threads (\textbf{\%})} & \textbf{\# chat excerpts (\textbf{\%})} \\ \hline
Help wanted in the context of \Actions{} (no error message) && 132 (28.82\%) & 7 (17.50\%) \\
Marginal mention of \Actions{}                              && 126 (27.51\%) & 14 (35.00\%) \\
Error/debug message in the context of \Actions{}            && 87 (19.00\%) & 2 (5.00\%) \\
Potential of using \Actions{}                               && 67 (14.63\%) & 14 (35.00\%) \\
Issue reproducing output with \Actions{}                    && 20 (4.37\%) &  0 (0\%) \\
Plan to use \Actions{}                                      && 16 (3.49\%) &  1 (2.50\%) \\
Non-English thread                                          && 7 (1.53\%) &  0 (0\%) \\
Other                                                       && 3 (0.66\%) &  2 (5.00\%) \\
\hline
\end{tabular}
\label{tab:issues}
\end{table}

\MyPara{Help wanted in the context of \Actions{} (no error message):} The largest group of Discussion threads that mention \Actions{} concerns requests for help in the context of the feature. We distinguish requests for help that mention a specific error message and are primarily aimed at soliciting help in debugging from those that are less specific. Conversations that do not provide a specific error message might ask for help in configuring a particular Action or mention that automation is not working as intended.

\MyPara{Marginal mention of \Actions{}:} While all threads and chat excerpts in our dataset contain the phrase `GitHub Action,' the feature is not the main topic of all such conversations. In some cases, \Actions{} is mentioned as part of a long discussion thread announcing a release where \Actions{} only affected a small number of features. In other cases, \Actions{} are only mentioned several months after the threads were started, and they are only marginally related to the thread topic.

\MyPara{Error/debug message in the context of \Actions{}:} Complementing the first category discussed above (Help wanted in the context of \Actions{}), Error/debug message in the context of \Actions{} contains discussions that start with a specific error or warning message and ask for help. In most cases, the error or warning has been provided verbatim by the developer starting the discussion. Errors can come from the \Actions{} feature itself or from the various applications, such as linters or code review bots, that are invoked via a \Actions{}.

\MyPara{Potential of using \Actions{}:} Since \Actions{} is still a relatively new feature, not all developers are aware of it. This category captures discussions in which developers suggest the use of \Actions{} to address a specific task, e.g., ``alternatively, the JIRA issue transitions at both PR creation and merge can be accomplished using \Actions{} listening to those events''\footnote{\url{https://github.com/cli/cli/discussions/3264}} or ``you could use the Vercel CLI directly as part of a GitHub Action (or similar) to deploy when releasing''.\footnote{\url{https://github.com/vercel/next.js/discussions/20905}}

\MyPara{Issue reproducing output with \Actions{}:} In many cases, the goal of using a GitHub Action is to automate a process otherwise conducted manually (or using a different tool). Discrepancies can occur when developers struggle to reproduce results they achieved with the help of a GitHub Action, e.g., ``This only happens with builds in \Actions{} and I am unable to reproduce this locally''.\footnote{\url{https://github.com/gatsbyjs/gatsby/discussions/32773}}

\MyPara{Plan to use \Actions{}:} Compared to the large number of GitHub issues dedicated to discussing projects' migration plans to \Actions{}, which we identified in our previous work, we found a smaller number of such discussion threads in this work, likely because the \Actions{} feature is more established now. An example of such a discussion thread is ``Migrating from Azure Pipelines to \Actions{}'',\footnote{\url{https://github.com/hyperledger/fabric/discussions/2456}} a thread that discusses the pros and cons of migration as well as how to implement it for a specific project.

\MyPara{Non-English thread:} A small number of discussion threads in our dataset were not in English.

\MyPara{Other:} Three of the discussion threads in our dataset did not fit any of the above categories and were assigned to the `Other' category. An example is a discussion thread on GitHub's docs project\footnote{\url{https://github.com/github/docs/discussions/2501}} about how to structure documentation about \Actions{}.

\rqtwo[
    \tcblower
    Discussion threads and chat excerpts that mention \Actions{} predominantly focus on requests for help in the context of the feature, with or without concrete error messages. A smaller group of discussions concerns plans for using the feature or debating its potential.
    \newline \newline
    \textbf{Comparison to our previous work}: A couple of years after the data collection for our previous work, in which we analyzed GitHub issues about \Actions{} (since GitHub Discussions did not yet exist), we now find fewer discussions about the potential of \Actions{} and more discussions about specific issues, such as errors and discrepancies.
]{}

\subsection{What is the Impact of \Actions{}? (RQ3)} 

To answer this question, we investigated the effects of GitHub Action adoption on project activities along four dimensions: (i) merged and non-merged pull requests, (ii) human conversation, (iii) efficiency to close pull requests, and (iv) modification effort. We start by investigating how Action adoption impacts the number of merged and non-merged pull requests. We fit two mixed-effect RDD models, as described in Section \ref{sec-statistical-modeling}. For these models, the \textit{number of merged/non-merged pull requests} per month is the dependent variable. Table~\ref{tab:resultspullrequest} summarizes the results of these models. In addition to the model coefficients, the table also shows the sum of squares, with variance explained for each variable. We also highlighted the time series predictors \textit{time}, \textit{time after intervention}, and \textit{intervention} in \textbf{bold}.

\begin{table}[htbp]
\scalefont{0.9}
\centering
\caption{The Effects of \Actions{} on PRs. The response is \textbf{log(number of merged/non-merged PRs)} per month.}
\label{tab:resultspullrequest}
\begin{threeparttable}
\begin{tabular}{lrrrrrr}
\midrule
 & \multicolumn{2}{c}{Merged PRs} & & \multicolumn{2}{c}{Non-merged PRs}\\
\cmidrule{2-3}\cmidrule{5-6}
 & Coeffs & Sum Sq. & & Coeffs & Sum Sq. \\
\cmidrule{2-3}\cmidrule{5-6}
Intercept & -0.603***  & & & -0.820*** &\\
TimeSinceFirstPR & -0.001** &  0.5 & & -0.002*** & 1.5  \\
log(TotalPRAuthors) & -0.054*** &  694.6 & & 0.136*** & 457.5 \\
log(TotalCommits) & 0.099*** & 206.7 & &  0.006 &  65.9 \\
log(OpenedPRs) & 0.841*** & 10688.9 & & 0.403*** & 3349.3 \\
log(PRComments) &  0.081*** & 240.0 & & 0.310*** & 1428.8 \\
log(PRCommits) & 0.270*** & 295.8 & & 0.151*** & 214.6 \\
\textbf{time} & -0.00003 & 0.0 & & 0.012*** &  8.3 \\
\textbf{interventionTrue} & 0.036*** & 2.2 & & -0.041*** & 2.9 \\
\textbf{time\_after\_intervention} & -0.008*** & 1.5 && -0.007** & 1.1 \\
\midrule
Marginal $R^2$ & 0.87 &  &  & 0.71 & \\
Conditional $R^2$ &  0.93 &  &  & 0.82 & \\
\midrule
\end{tabular}
\begin{tablenotes}
 \item *** $p < 0.001$, ** $p < 0.01$, * $p < 0.05$
 \item Time series predictors in \textbf{bold}.
\end{tablenotes}
\end{threeparttable}
\end{table}

Analyzing the model for merged pull requests, we found that the fixed-effects part fits the data well ($R^2_m=0.87$). However, considering $R^2_c=0.93$, variability also appears from project-to-project and language-to-language. Among the fixed effects, we note that the number of monthly pull requests explains most of the variability in the model, indicating that projects receiving more contributions tend to have more merged pull requests, with other variables held constant. Regarding the Action effects, there is a discontinuity at adoption time, followed by a statistically significant decrease after the introduction.

Similar to the previous model, the fixed-effect part of the non-merged pull requests model fits the data well ($R^2_m=0.71$), even though a considerable amount of variability is explained by random effects ($R^2_c=0.82$). We  note  similar results on fixed effects: projects receiving more contributions tend to have more non-merged pull requests. In addition, pull requests receiving more comments tend to be rejected. The effect of Action adoption on the non-merged pull requests differs from the previous model. Regarding the time series predictors, the negative trend in the number of non-merged pull requests before the Action adoption is reversed, toward an increase after adoption.

\begin{table}[htbp]
\scalefont{0.9}
\centering
\caption{The Effects of \Actions{} on Pull Request Comments. The response is \textbf{log(median of comments)} per month.}
\label{tab:resultscomments}
\begin{threeparttable}
\begin{tabular}{lrrrrrr}
\midrule
 & \multicolumn{2}{c}{Merged PRs} & & \multicolumn{2}{c}{Non-merged PRs}\\
\cmidrule{2-3}\cmidrule{5-6}
 & Coeffs & Sum Sq. & & Coeffs & Sum Sq. \\
\cmidrule{2-3}\cmidrule{5-6}
Intercept &  -0.086 & & & -0.196*** &\\
TimeSinceFirstPR & -0.0004 & 0.28 & & 0.0002 & 5.7 \\
log(TotalPRAuthors) & 0.053*** & 14.52 & & 0.002 & 110.1 \\
log(TotalCommits) & -0.011 & 0.12 & & 0.028*** & 22.9 \\
log(OpenedPRs) & -0.013*** & 31.04 &  & 0.083*** & 498.4 \\
log(TimeToClosePRs) & 0.066*** & 1258.88 & &  0.108*** & 3828.9 \\
log(PRCommits) &  0.355*** & 497.59 & & 0.215*** & 461.9 \\
\textbf{time} & -0.001 & 0.10 & & 0.0002 & 11.2 \\
\textbf{interventionTrue} & -0.023*** & 0.90 & & 0.034*** & 2.0 \\
\textbf{time\_after\_intervention} & 0.006*** &  0.92 && -0.018*** & 7.9 \\
\midrule
Marginal $R^2$ & 0.30 & & &  0.56 & \\
Conditional $R^2$ & 0.58 & & & 0.69 & \\
\midrule
\end{tabular}
\begin{tablenotes}
 \item *** $p < 0.001$, ** $p < 0.01$, * $p < 0.05$
 \item Time series predictors in \textbf{bold}.
\end{tablenotes}
\end{threeparttable}
\end{table}

To investigate the effects of Action adoption on pull request communication, we fit one model to merged pull requests and another to non-merged ones. The \textit{median of pull request comments} per month is the dependent variable. Table~\ref{tab:resultscomments} shows the results of the fitted models. Considering the model of comments on merged pull requests, we found that the combined fixed-and-random effects ($R^2_c=0.58$) fit the data better than the fixed effects ($R^2_m=0.30$), showing that most of the explained variability in the data is associated with project-to-project and language-to-language variability, rather than the fixed effects. We also observe that the time to close pull requests explains the largest amount of variability in the model, indicating that the communication during the pull request review is strongly associated with the time to merge it. Regarding the Action effects, we note no statistically significant trend before adoption; a discontinuity at the adoption time; and an apparent increase in the time trend after adoption.

Turning to the model of comments on non-merged pull requests, the model fits the data well ($R^2_m=0.56$), and variability is explained by the random variables ($R^2_c=0.69$). This model also suggests that communication during the pull request review is strongly associated with the time to reject the pull request. Table~\ref{tab:resultscomments} shows a discontinuity at adoption time, followed by a statistically significant decrease after Action adoption.

\begin{table}[htbp]
\scalefont{0.9}
\centering
\caption{The effects of \Actions{} on time to close PRs. The response is \textbf{log(median of time to close PRs)} per month.}
\label{tab:resultstime}
\begin{threeparttable}
\begin{tabular}{lrrrrrr}
\midrule
 & \multicolumn{2}{c}{Merged PRs} & & \multicolumn{2}{c}{Non-merged PRs}\\
\cmidrule{2-3}\cmidrule{5-6}
 & Coeffs & Sum Sq. & & Coeffs & Sum Sq. \\
\cmidrule{2-3}\cmidrule{5-6}
Intercept & -0.803*** & & &  -0.064 &\\
TimeSinceFirstPR & 0.002 & 51.6 & &  0.001 & 158  \\
log(TotalPRAuthors) &  0.357*** & 373.3 & & 0.291*** & 2075 \\
log(TotalCommits) &  -0.037 & 0.1 & & -0.133*** & 145 \\
log(OpenedPRs) &  -0.216*** & 135.3 & & -0.059*** & 5211 \\
log(PRComments) &  1.262*** & 24644.5 & & 2.763*** & 93257 \\
log(PRCommits) & 1.698*** & 11716.7 & & 0.698*** & 4617 \\
\textbf{time} & -0.0003 & 13.2 & & 0.039*** &  57 \\
\textbf{interventionTrue} & -0.124*** & 26.1 & & -0.327*** & 181 \\
\textbf{time\_after\_intervention} & 0.019** & 9.3 && 0.019 &  9 \\
\midrule
Marginal $R^2$ & 0.35 &  & & 0.50 & \\
Conditional $R^2$ &  0.55 &  & & 0.61 & \\
\midrule
\end{tabular}
\begin{tablenotes}
 \item *** $p < 0.001$, ** $p < 0.01$, * $p < 0.05$
 \item Time series predictors in \textbf{bold}.
\end{tablenotes}
\end{threeparttable}
\end{table}

We fitted two RDD models where \textit{median of time to close pull requests} per month is the dependent variable. The results are shown in Table~\ref{tab:resultstime}. Analyzing the results of the effect of \Actions{} on the latency to merge pull requests, we found that combined fixed-and-random effects fit the data better than the fixed effects. Although several variables affect the trends of pull request latency, communication during the pull requests is responsible for most of the variability in the data. This indicates the expected results: the more effort contributors expend discussing the contribution, the more time the contribution takes to merge. The number of commits also explains the amount of data variability, since a project with many changes needs more time to review and merge them. We observe a discontinuity at adoption time, followed by a statistically significant decrease after GitHub Action's introduction.

Turning to the model of non-merged pull requests, we note that it fits the data well ($R^2_m=0.50$), and variability is explained by the random variables ($R^2_c=0.61$). As above, communication during the pull requests is responsible for most of the variability encountered in the results. Similar to the previous model, none of the Action-related predictors have statistically significant effects on the time to reject pull requests. We observe an increasing trend before adoption, followed by a statistically significant discontinuity at adoption. After adoption, however, there is no effect on the time to reject pull requests, since the time after intervention coefficient is not statistically significant.

\begin{table}[htbp]
\scalefont{0.9}
\centering
\caption{The Effects of \Actions{} on Pull Request Commits. The response is \textbf{log(median of commits)} per month.}
\label{tab:resultscommits}
\begin{threeparttable}
\begin{tabular}{lrrrrrr}
\midrule
 & \multicolumn{2}{c}{Merged PRs} & & \multicolumn{2}{c}{Non-merged PRs}\\
\cmidrule{2-3}\cmidrule{5-6}
 & Coeffs & Sum Sq. & & Coeffs & Sum Sq. \\
\cmidrule{2-3}\cmidrule{5-6}
Intercept &  0.342*** & && 0.235*** &\\ 
TimeSinceFirstPR & -0.0001 & 0.00 & & 0.0003 & 2.28 \\
log(TotalPRAuthors) & -0.041*** & 28.90 & & -0.042*** & 111.58 \\
log(TotalCommits) & 0.010 & 7.11 & & 0.008 & 37.73 \\
log(OpenedPRs) & 0.130*** & 327.96 && 0.101*** & 573.78 \\
log(PRComments) & 0.444*** & 1189.16 && 0.524*** & 2625.12 \\
\textbf{time} & 0.002 & 3.72 && -0.006*** & 0.11 \\
\textbf{interventionTrue} & 0.036*** & 2.14 & & 0.004 & 0.03 \\
\textbf{time\_after\_intervention} & -0.006*** & 1.01 && 0.013*** & 4.15 \\
\midrule
Marginal $R^2$ & 0.37 &  & & 0.36 & \\
Conditional $R^2$ & 0.60 &  & & 0.48 & \\
\midrule
\end{tabular}
\begin{tablenotes}
 \item *** $p < 0.001$, ** $p < 0.01$, * $p < 0.05$
 \item Time series predictors in \textbf{bold}.
\end{tablenotes}
\end{threeparttable}
\end{table}

Finally, we studied whether Action adoption affects the number of commits made before and during the pull request review. Again, we fitted two models for merged and non-merged pull requests, where the \textit{median of pull request commits} per month is the dependent variable. The results are shown in Table~\ref{tab:resultscommits}. Analyzing the model of commits on merged pull requests, we found that the combined fixed-and-random effects ($R^2_c=0.60$) fit the data better than the fixed effects ($R^2_m=0.37$). The statistical significance of all Action-related coefficients indicates that the adoption of \Actions{} affected the number of commits. We note a statistically significant discontinuity at adoption time, followed by a decreasing trend after adoption. Additionally, we can also observe that the number of pull request comments and the number of contributions per month explains most of the variability in the result. This result suggests that the more comments and pull requests there are, the more commits there will be. 

Investigating the results of the non-merged pull request model, we also found that the combined fixed-and-random effects fit the data better than the fixed effects. Similar to the previous model, the number of pull request comments per month explains most of the results' variability. Regarding the time series predictors, the model did not detect any discontinuity at adoption time. However, the negative trend in the median of commits before the bot adoption is reversed, toward an increase after adoption.

\rqthree[
    \tcblower
    After adopting \Actions{}, on average, there are fewer accepted pull requests, with more discussion comments and fewer commits, which take more time to merge. On the other hand, there are more rejected pull requests, which contain fewer comments and more commits.
    \newline \newline
    \textbf{Comparison to our previous work}: We confirm the results from our previous work. We have already shown that \Actions{} increase the number of rejected pull requests and decrease the number of commits on merged pull requests.
    ]{}

\subsection{How does the impact of \Actions{} differ across Action categories? (RQ4)}

To investigate the effects of GitHub Action adoption on project activities across the four most used Action categories in our dataset, we fit thirty-two mixed-effect RDD models, as described in Section \ref{sec-statistical-modeling}. We considered the same activity indicators studied in the previous research question: (i) merged and non-merged pull requests, (ii) human conversation, (iii) efficiency to close pull requests, and (iv) modification effort.

\begin{table}[htbp]
\scalefont{0.7}
\centering
\caption{The Effects of \Actions{} on Merged Pull Requests. The response is \textbf{log(number of merged PRs)} per month.}
\label{tab:mergedprscat}
\begin{threeparttable}
\begin{tabular}{l|r|r|r|r}
\midrule
 & \multicolumn{1}{c}{Utilities} & \multicolumn{1}{c}{CI}  & \multicolumn{1}{c}{Code Quality} & \multicolumn{1}{c}{Deployment}\\
\midrule
 & Coeffs (SS) & Coeffs (SS) & Coeffs (SS) & Coeffs (SS) \\
\midrule
Intercept &  -0.711***  & -0.938*** & -0.752*** & -0.769***\\
TimeSinceFirstPR &  -0.002*** (0.1) & -0.003*** (2.76)  & -0.0004 (4.61) & -0.002** (14.958)\\
log(TotalPRAuthors) & -0.030* (845.7) & 0.037 (372.20) & -0.071** (262.96) & -0.080** (216.624) \\
log(TotalCommits) &  0.105*** (243.4) & 0.097*** (83.95) & 0.093*** (88.59) & 0.154*** (70.461)\\
log(OpenedPRs) & 0.823*** (6074.3) & 0.838*** (727.16) & 0.901*** (827.66) & 0.876*** (294.845)\\
log(PRComments) &  0.075*** (152.4) & 0.066*** (14.34) & 0.125*** (18.11) & 0.065* (2.123)\\
log(PRCommits) & 0.285*** (201.9) & 0.287*** (20.65) & 0.246*** (13.58) & 0.226*** (4.602)\\
\textbf{time} & 0.001 (0.3) & 0.002 (0.02) & -0.015* (0.00) & -0.013 (0.003)\\
\textbf{interventionTrue} &  0.035*** (1.2) & 0.029 (0.09) & 0.063 (0.55) & 0.122*** (0.685)\\
\textbf{time\_after\_intervention} & -0.008** (1.0) & -0.012 (0.27) & 0.013 (0.24) & -0.005 (0.013)\\
\midrule
Marginal $R^2$ & 0.88 & 0.91 & 0.70 & 0.95\\
Conditional $R^2$ &  0.93 & 0.94 & 0.92 & 0.96\\
\midrule
\end{tabular}
\begin{tablenotes}
 \item *** $p < 0.001$, ** $p < 0.01$, * $p < 0.05$
 \item SS stands for ``Sum of Squares''
 \item Time series predictors in \textbf{bold}.
\end{tablenotes}
\end{threeparttable}
\end{table}

\begin{table}[htbp]
\scalefont{0.7}
\centering
\caption{The Effects of \Actions{} on Non-merged Pull Requests. The response is \textbf{log(number of non-merged PRs)} per month.}
\label{tab:nonmergedprscat}
\begin{threeparttable}
\begin{tabular}{l|r|r|r|r}
\midrule
 & \multicolumn{1}{c}{Utilities} & \multicolumn{1}{c}{CI}  & \multicolumn{1}{c}{Code Quality} & \multicolumn{1}{c}{Deployment}\\
\midrule
 & Coeffs (SS) & Coeffs (SS) & Coeffs (SS) & Coeffs (SS) \\
\midrule
Intercept &  -0.744*** & -0.941*** & -1.281*** & -1.382***\\
TimeSinceFirstPR &  -0.002*** (0.77) & -0.002** (0.955) & -0.004*** (0.449) & -0.001 (2.864)\\
log(TotalPRAuthors) & 0.127*** (450.75) & 0.154*** (125.377) & 0.229*** (89.943) & 0.191** (26.420)\\
log(TotalCommits) &  0.004 (66.37) & 0.008 (16.053) & 0.010 (12.975) & 0.062 (3.497)\\
log(OpenedPRs) & 0.400*** (1941.50) & 0.413*** (238.575) & 0.387*** (208.773) & 0.329*** (46.801)\\
log(PRComments) &  0.318*** (846.53) & 0.327*** (104.532) & 0.312*** (98.287) & 0.196*** (14.421)\\
log(PRCommits) & 0.153*** (127.53) & 0.124*** (10.735) & 0.152*** (11.747) & 0.182*** (7.644)\\
\textbf{time} & 0.009*** (2.95) & 0.008 (0.351) & 0.016* (0.069) & 0.033*** (0.645)\\
\textbf{interventionTrue} & -0.038*** (1.87) & -0.032 (0.179) & 0.004 (0.008) & -0.112* (0.708) \\
\textbf{time\_after\_intervention} & -0.002 (0.05) & -0.001 (0.001) & -0.026** (0.893) & -0.019 (0.204)\\
\midrule
Marginal $R^2$ & 0.71 & 0.74 & 0.67 & 0.72\\
Conditional $R^2$ &  0.82 & 0.84 & 0.79 & 0.86\\
\midrule
\end{tabular}
\begin{tablenotes}
 \item *** $p < 0.001$, ** $p < 0.01$, * $p < 0.05$
 \item SS stands for ``Sum of Squares''
 \item Time series predictors in \textbf{bold}.
\end{tablenotes}
\end{threeparttable}
\end{table}

We fitted four RDD models for each of the Action categories where \textit{number of merged pull requests} per month is the dependent variable. The results are shown in Table~\ref{tab:mergedprscat}. The statistical significance of the time series predictors for utilities indicates that the adoption of \Actions{} of this category affected the trend in the number of merged pull requests. In addition, we fitted four RDD models where \textit{number of non-merged pull requests} per month is the dependent variable (see Table~\ref{tab:nonmergedprscat}). In the model of code quality \Actions{}, although the model did not detect any discontinuity at adoption time, the positive trend in the number of rejected pull requests before Action adoption is reversed toward a decrease after adoption. Considering the other categories, the Action-related predictors do not have statistically significant effects, meaning the trend in the number of merged and non-merged pull requests is stationary over time and remains unaffected by the Action adoption.

\begin{table}[htbp]
\scalefont{0.7}
\centering
\caption{The Effects of \Actions{} on Comments of Merged Pull Requests. The response is \textbf{log(number of comments on merged PRs)} per month.}
\label{tab:mergedcommentscat}
\begin{threeparttable}
\begin{tabular}{l|r|r|r|r}
\midrule
 & \multicolumn{1}{c}{Utilities} & \multicolumn{1}{c}{CI}  & \multicolumn{1}{c}{Code Quality} & \multicolumn{1}{c}{Deployment}\\
\midrule
 & Coeffs (SS) & Coeffs (SS) & Coeffs (SS) & Coeffs (SS) \\
\midrule
Intercept &  -0.114  & 0.018 & 0.340* & -0.257\\
TimeSinceFirstPR &  -0.0003  (0.42) & -0.0004 (0.134) & 0.0002 (0.002) & -0.001 (0.1381)\\
log(TotalPRAuthors) & 0.053*** (14.26) & 0.082** (2.572) & -0.005 (0.329) & 0.133** (1.0812)\\
log(TotalCommits) &  -0.006 (0.21) & -0.047* (0.009) & -0.041 (0.024) & -0.019 (0.0018)\\
log(OpenedPRs) & -0.018*** (18.63) & -0.0001 (3.828) & 0.008 (4.016) & -0.019 (0.0477)\\
log(TimeToClosePRs) & 0.068*** (803.26) & 0.058*** (75.501) & 0.067*** (66.694) & 0.044*** (9.8729)\\
log(PRCommits) &  0.349*** (289.59) & 0.421*** (43.721) & 0.356*** (28.00) & 0.279*** (5.82)\\
\textbf{time} & -0.002 (0.02) & -0.006 (0.093) & 0.001 (0.703) & -0.013 (0.473)\\
\textbf{interventionTrue} & -0.021** (0.38) & -0.004 (0.00) & -0.002 (0.001) & 0.031 (0.051)\\
\textbf{time\_after\_intervention} & 0.007** (0.79) & 0.009 (0.173) & 0.010 (0.125) & 0.004 (0.008)\\
\midrule
Marginal $R^2$ & 0.30 & 0.27 & 0.27 & 0.28\\
Conditional $R^2$ & 0.58 & 0.61 & 0.54 & 0.74\\
\midrule
\end{tabular}
\begin{tablenotes}
 \item *** $p < 0.001$, ** $p < 0.01$, * $p < 0.05$
 \item SS stands for ``Sum of Squares''
 \item Time series predictors in \textbf{bold}.
\end{tablenotes}
\end{threeparttable}
\end{table}

\begin{table}[htbp]
\scalefont{0.7}
\centering
\caption{The Effects of \Actions{} on Comments of Non-merged Pull Requests. The response is \textbf{log(number of comments on non-merged PRs)} per month.}
\label{tab:nonmergedcommentscat}
\begin{threeparttable}
\begin{tabular}{l|r|r|r|r}
\midrule
 & \multicolumn{1}{c}{Utilities} & \multicolumn{1}{c}{CI}  & \multicolumn{1}{c}{Code Quality} & \multicolumn{1}{c}{Deployment}\\
\midrule
 & Coeffs (SS) & Coeffs (SS) & Coeffs (SS) & Coeffs (SS) \\
\midrule
Intercept &  -0.211***  & -0.013 & -0.025 & -0.355\\
TimeSinceFirstPR &  0.0002 (5.37) & 0.0002 (2.102) & 0.001 (2.904) & 0.0002 (1.248)\\
log(TotalPRAuthors) & -0.003 (105.70) & 0.023 (21.696) & -0.050 (11.326) & 0.043 (6.091)\\
log(TotalCommits) &  0.033*** (23.66) & -0.021 (3.051) & 0.035 (3.487) & 0.017 (0.880)\\
log(OpenedPRs) & 0.081*** (292.09) & 0.120*** (44.057) & 0.062*** (35.652) & 0.104*** (7.606) \\
log(TimeToClosePRs) &  0.107*** (2163.48) & 0.106*** (276.011) & 0.118*** (284.727) & 0.101*** (59.142)\\
log(PRCommits) &  0.204*** (236.52) & 0.199*** (30.276) & 0.246*** (33.097) & 0.141*** (5.055)\\
\textbf{time} & 0.001 (5.94) & 0.009 (0.196) & -0.009 (0.146) & 0.006 (0.882)\\
\textbf{interventionTrue} & 0.037*** (0.92) & 0.029 (0.040) & 0.061 (0.402) & 0.007 (0.003)\\
\textbf{time\_after\_intervention} & -0.018*** (4.94) & -0.027*** (1.512) & -0.003 (0.013) & -0.030** (0.522)\\
\midrule
Marginal $R^2$ & 0.56 & 0.54 & 0.53 & 0.57\\
Conditional $R^2$ & 0.68 & 0.68 & 0.63 & 0.76\\
\midrule
\end{tabular}
\begin{tablenotes}
 \item *** $p < 0.001$, ** $p < 0.01$, * $p < 0.05$
 \item SS stands for ``Sum of Squares''
 \item Time series predictors in \textbf{bold}.
\end{tablenotes}
\end{threeparttable}
\end{table}

Analyzing the models of human discussions (see Table~\ref{tab:mergedcommentscat}), where the \textit{median of comments per month in merged pull requests} is the dependent variable, we found that the introduction of utility \Actions{} increases the discussions by developers on merged pull requests. There is a discontinuity at adoption time, followed by a statistically significant decrease after the utilities' introduction. Turning to the models where the \textit{median of comments per month in rejected pull requests} is the dependent variable, we found that utilities, CI, and deployment \Actions{} decreased the number of comments on rejected pull requests.

\begin{table}[htbp]
\scalefont{0.7}
\centering
\caption{The effects of \Actions{} on the time to merge pull requests. The response is \textbf{log(median of time to merge PRs)} per month.}
\label{tab:mergedtime}
\begin{threeparttable}
\begin{tabular}{l|r|r|r|r}
\midrule
 & \multicolumn{1}{c}{Utilities} & \multicolumn{1}{c}{CI}  & \multicolumn{1}{c}{Code Quality} & \multicolumn{1}{c}{Deployment}\\
\midrule
 & Coeffs (SS) & Coeffs (SS) & Coeffs (SS) & Coeffs (SS) \\
\midrule
Intercept & -0.672*** & -1.664*** & -1.078 & -2.200**\\
TimeSinceFirstPR & 0.002 (50.4) & 0.003 (26.37) & 0.001 (6.83) & 0.006 (45.16)\\
log(TotalPRAuthors) & 0.315**** (368.7) & 0.307** (53.63) & 0.480*** (39.51) & 0.449** (59.82)\\
log(TotalCommits) & -0.054 (0.8) & 0.105 (1.93) & -0.071 (1.87) & 0.108 (0.55)\\
log(OpenedPRs) & -0.183*** (202.2) & -0.327*** (0.09) & -0.235*** (4.28) & -0.384*** (3.81)\\
log(PRComments) & 1.294*** (15948.1) & 1.151*** (1500.36) & 1.268*** (1288.99) & 1.384*** (322.79)\\
log(PRCommits) & 1.739*** (7522.1) & 1.723*** (737.90) & 1.608*** (575.56) & 1.442*** (172.35)\\
\textbf{time} & 0.015* (0.4) & 0.0003 (0.38) & -0.010 (5.09) & 0.005 (2.02)\\
\textbf{interventionTrue} & -0.183*** (38.6) & -0.107 (1.63) & -0.009 (0.02) & 0.008 (0.01)\\
\textbf{time\_after\_intervention} & 0.012 (2.2) & 0.017 (0.59) & -0.008 (0.08) & 0.016 (0.14)\\
\midrule
Marginal $R^2$ & 0.36 & 0.29 & 0.31 & 0.36\\
Conditional $R^2$ & 0.55 & 0.52 & 0.52 & 0.52\\
\midrule
\end{tabular}
\begin{tablenotes}
 \item *** $p < 0.001$, ** $p < 0.01$, * $p < 0.05$
 \item SS stands for ``Sum of Squares''
 \item Time series predictors in \textbf{bold}.
\end{tablenotes}
\end{threeparttable}
\end{table}

\begin{table}[htbp]
\scalefont{0.7}
\centering
\caption{The effects of \Actions{}{} on the time to close pull requests. The response is \textbf{log(median of time to close PRs)} per month.}
\label{tab:nonmergedtime}
\begin{threeparttable}
\begin{tabular}{l|r|r|r|r}
\midrule
 & \multicolumn{1}{c}{Utilities} & \multicolumn{1}{c}{CI}  & \multicolumn{1}{c}{Code Quality} & \multicolumn{1}{c}{Deployment}\\
\midrule
 & Coeffs (SS) & Coeffs (SS) & Coeffs (SS) & Coeffs (SS) \\
\midrule
Intercept &  0.023  & -0.895* & -0.177 & 0.038\\
TimeSinceFirstPR &  0.001 (147.0) & 0.005 (108.7) & 0.001 (63.5) & 0.004 (39.80)\\
log(TotalPRAuthors) & 0.309*** (2213.0) & 0.118 (346.3) & 0.509*** (348.4) & 0.094 (98.06)\\
log(TotalCommits) &  -0.160*** (151.0) & 0.120 (60.9) & -0.285** (3.8) & -0.054 (9.90)\\
log(OpenedPRs) & -0.039* (3436.0) & -0.241*** (246.7) & 0.004 (431.1) & -0.160 (66.34)\\
log(PRComments) &  2.757*** (54150.0) & 2.795*** (7146.2) & 2.642*** (6122.8) & 2.827*** (1687.45)\\
log(PRCommits) & 0.772*** (3283.0) & 0.681*** (337.8) & 0.742*** (291.0) &  0.797*** (153.57)\\
\textbf{time} & 0.031*** (41.0) & 0.011 (1.1) & 0.075* (0.0) & 0.128** (24.97)\\
\textbf{interventionTrue} & -0.280*** (86.0) & -0.226 (7.9) & -0.410** (20.9) & -0.650** (20.03)\\
\textbf{time\_after\_intervention} & 0.031* (15.0) & 0.020 (0.8) & -0.040 (2.2) & 0.010 (0.06)\\
\midrule
Marginal $R^2$ & 0.50 & 0.47 & 0.50 & 0.49\\
Conditional $R^2$ & 0.61 & 0.58 & 0.58 & 0.63\\
\midrule
\end{tabular}
\begin{tablenotes}
 \item *** $p < 0.001$, ** $p < 0.01$, * $p < 0.05$
 \item SS stands for ``Sum of Squares''
 \item Time series predictors in \textbf{bold}.
\end{tablenotes}
\end{threeparttable}
\end{table}

\begin{table}[htbp]
\scalefont{0.7}
\centering
\caption{The Effects of \Actions{}{} on Commits of Merged Pull Requests. The response is \textbf{log(number of commit on merged PRs)} per month.}
\label{tab:mergedcommits}
\begin{threeparttable}
\begin{tabular}{l|r|r|r|r}
\midrule
 & \multicolumn{1}{c}{Utilities} & \multicolumn{1}{c}{CI}  & \multicolumn{1}{c}{Code Quality} & \multicolumn{1}{c}{Deployment}\\
\midrule
 & Coeffs (SS) & Coeffs (SS) & Coeffs (SS) & Coeffs (SS) \\
\midrule
Intercept & 0.360*** & 0.388*** & 0.284* & 0.375\\
TimeSinceFirstPR & -0.0001 (0.02) & 0.001 (0.335) & -0.0002 (0.000) & 0.0004 (0.1896)\\
log(TotalPRAuthors) & -0.042*** (32.11) & -0.067** (4.466) & -0.034 (2.671) & -0.088 (1.5858)\\
log(TotalCommits) & 0.008 (7.76) & 0.015 (1.270) & 0.026 (1.139) & 0.032 (0.5889)\\
log(OpenedPRs) & 0.138*** (213.85) & 0.128*** (22.456) & 0.113*** (18.121) & 0.153*** (7.9283)\\
log(PRComments) & 0.460*** (764.69) & 0.403*** (70.617) & 0.420*** (60.073) & 0.425*** (13.8455)\\
\textbf{time} & 0.0004 (1.18) & 0.005 (0.531) & 0.005 (0.091) & -0.003 (0.0398)\\
\textbf{interventionTrue} & 0.040*** (1.74) & 0.025 (0.079) & -0.001 (0.002) & 0.067 (0.1982)\\
\textbf{time\_after\_intervention} & -0.006** (0.49) & -0.007 (0.110) & -0.006 (0.042) & -0.007 (0.026)\\
\midrule
Marginal $R^2$ & 0.38 & 0.37 & 0.30 & 0.35\\
Conditional $R^2$ & 0.59 & 0.62 & 0.55 & 0.69\\
\midrule
\end{tabular}
\begin{tablenotes}
 \item *** $p < 0.001$, ** $p < 0.01$, * $p < 0.05$
 \item SS stands for ``Sum of Squares''
 \item Time series predictors in \textbf{bold}.
\end{tablenotes}
\end{threeparttable}
\end{table}

\begin{table}[htbp]
\scalefont{0.7}
\centering
\caption{The Effects of \Actions{}{} on Commits of Non-merged Pull Requests. The response is \textbf{log(number of commit on non-merged PRs)} per month.}
\label{tab:nonmergedcommits}
\begin{threeparttable}
\begin{tabular}{l|r|r|r|r}
\midrule
 & \multicolumn{1}{c}{Utilities} & \multicolumn{1}{c}{CI}  & \multicolumn{1}{c}{Code Quality} & \multicolumn{1}{c}{Deployment}\\
\midrule
 & Coeffs (SS) & Coeffs (SS) & Coeffs (SS) & Coeffs (SS) \\
\midrule
Intercept &  0.196*** & 0.457*** & 0.270 & 0.344\\
TimeSinceFirstPR &  0.0003 (2.11) & 0.001 (1.163) & 0.0004 (1.105) & -0.0005 (0.689)\\
log(TotalPRAuthors) & -0.035** (113.35) & -0.084** (17.388) & -0.049 (12.732) & -0.032 (5.763)\\
log(TotalCommits) &  0.010 (37.98) & 0.008 (6.760) & 0.003 (6.040) & 0.012 (1.455)\\
log(OpenedPRs) & 0.101*** (341.01) & 0.118*** (47.491) & 0.119*** (44.609) & 0.064** (7.677)\\
log(PRComments) & 0.517*** (1462.07) & 0.467*** (162.636) & 0.486*** (161.848) & 0.586*** (53.57) \\
\textbf{time} & -0.003 (0.01) & -0.014* (0.64) & 0.011 (0.089) & -0.020 (0.45)\\
\textbf{interventionTrue} & -0.011 (0.06) & 0.069* (0.81) & -0.016 (0.07) & -0.057 (0.07)\\
\textbf{time\_after\_intervention} & 0.009** (1.12) & 0.001 (0.001) & -0.020 (0.51) & 0.036 (0.74)\\
\midrule
Marginal $R^2$ & 0.36 & 0.32 & 0.34 & 0.29\\
Conditional $R^2$ & 0.48 & 0.45 & 0.44 & 0.50\\
\midrule
\end{tabular}
\begin{tablenotes}
 \item *** $p < 0.001$, ** $p < 0.01$, * $p < 0.05$
 \item SS stands for ``Sum of Squares''
 \item Time series predictors in \textbf{bold}.
\end{tablenotes}
\end{threeparttable}
\end{table}

\begin{table}[]
\caption{Segmented analysis comparison (whole sample vs. different categories)}
\label{tab:discussion}
\begin{tabular}{lcllll}
Indicator                & \multicolumn{1}{l}{Whole Sample} & Utilities             & CI                    & Code Quality          & Deployment            \\
Merged PRs             & -                                & \multicolumn{1}{c}{-} &                       &                       &                       \\
Non-merged PRs             & +                                &                       &                       & \multicolumn{1}{c}{-} &                       \\
Comments in Merged PRs & +                                & \multicolumn{1}{c}{+} &                       &                       &                       \\
Comments in Non-merged PRs & -                                & \multicolumn{1}{c}{-} & \multicolumn{1}{c}{-} &                       & \multicolumn{1}{c}{-} \\
Commits in Merged PRs  & -                                & \multicolumn{1}{c}{-} &                       &                       &                       \\
Commits in Non-merged PRs  & +                                & \multicolumn{1}{c}{+} &                       &                       &                       \\
Time to merge a PR      & +                                &                       &                       &                       &                       \\
Time to close a PR      & \multicolumn{1}{l}{}             &                       &                       &                       &                       \\                   
\end{tabular}
\end{table}

Segmenting the analysis for specific categories, we found that the number of comments in rejected pull requests statistically decreases in 3 out of 4 categories as well as in the whole sample, as can be observed in Table~\ref{tab:discussion}. Besides this indicator, the Utilities category, which contains the largest number of \Actions{}, resembles the whole sample and also showed statistical differences in accepted pull requests (decreased), comments in accepted pull requests (increased), commits in accepted pull requests (decreased), and commits in rejected pull requests (increased). In the Code Quality category, the only indicator for which we observed a statistically significant change is the number of rejected pull requests (decreased), which is in the opposite direction of the whole sample. We conjecture that Code Quality \Actions{} help contributors improve the quality of pull requests that would otherwise be rejected and, thus, the number of rejected pull requests in such repositories tends to decrease after the introduction of the Action.

\rqfour[
    \tcblower
    Analyzing the four most used types of \Actions{}, we found that the number of comments in rejected pull requests consistently decreased across categories (3 out of 4). Several other indicators also changed after the adoption of \Actions{} from the Utilities category: accepted pull requests (decreased), comments in accepted pull requests (increased), commits in accepted pull requests (decreased), and commits in rejected pull requests (increased). In the Code Quality category, the only indicator that changed is the number of rejected pull requests (decreased).
    ]{}

\section{Discussion}
This section discusses our results and the key implications for practitioners, researchers, and educators.

\textbf{Automation in Software Engineering.} The rise of \Actions{} evidence the importance of automation in software engineering. OSS project maintainers, who are often busy with coding and community-building activities, can save a lot of time by using \Actions{} to automate repetitive tasks such as replacing strings and running the integration pipeline. Automation can bring not only time savings but also avoid human errors and provide consistency in completed tasks~\cite{Storey2016}. The multiple benefits of automation can help explain the widespread adoption of \Actions{}. Indeed, we have seen an increase from 0.7\% to circa 30\% in the adoption of \Actions{} since we conducted our prior work~\cite{kinsman2021}. This result is in line with the studies conducted by Decan et al. \cite{decan22actions} and Chen et al.~\cite{chen2021let}, who found \Actions{} in 43.9\% and 22\% of their sample of projects, respectively. We also found a large number of projects discussing using \Actions{}. Given this impetus to automation, other software engineering tools and platforms should consider offering automation capabilities or integration endpoints and APIs so that the variety of tools used in software development can be integrated into large and more complex workflows. Our results show that projects have a median of four \Actions{}, and we expect this number to grow as more tools are integrated into the workflow pipelines. The power of automating tasks with \Actions{} can also be explored in other contexts. For example, software engineering educators can use \Actions{} to build automation tools to better support their assignments, including those related to contributing to OSS~\cite{pinto2017training}. \Actions{} can also automate multiple aspects related to code quality checking still unexplored~\cite{aniche2016satt,dos2018impacts,aniche2016validated}.

\textbf{Problems may arise from the integration of distinct automation tools.} We identified almost 1,000 distinct \Actions{} in the repositories, and projects often use more than one \Actions{} in their repositories (median number of four and a maximum of 46 in a single repository). Wessel et al. \cite{wessel2021don} showed that the use of multiple automation tools may cause noise and inconsistencies. As some \Actions{} provide limited configuration options and are hard to change, researchers and practitioners should find ways to seamlessly integrate such tools in their repositories. A promising approach is the use of meta-bots to integrate and moderate the interactions of multiple bots~\cite{wessel2022icse}. Such meta-bots can be responsible for mediating the communication between the tools and the environment. Another approach is the adoption of process execution languages, such as BPEL and BPMN~\cite{ouyang2006bpmn}, to allow end users to describe their workflow and how information moves among the activities, which may include manual and automated tasks. Future work can investigate how to facilitate such end-user programming to build complex workflows and automation scenarios. Approaches such as orchestrations and choreographies \cite{leite2013systematic} can also be investigated in this context. Future work can also investigate the interplay of \Actions{} and other automation tools, such as development bots~\cite{Wessel2018}. It is still not clear when each platform should be used and how the interoperability problems should be addressed.

\textbf{CI/CD is one of the most automated parts of the workflow.} Our results are in line with Golzadeh et al.~\cite{golzadehrise}, who showed that \Actions{} are replacing other continuous integration platforms. Almost one-fourth of the \Actions{} we found are categorized as continuous integration and many other categories of actions are closely related to continuous integration or continuous delivery, including deployment, publishing, testing, etc. The popularity of these types of \Actions{} can be explained by the popularity of CI/CD automation tools themselves. The literature has shown that these tools streamline the review of external contributions \cite{cassee2020silent}. Hilton et al. \cite{Hilton2016} showed that projects can process more outside contributions after the adoption of CI without any change in code quality. With less time spent reviewing external pull requests, maintainers can focus on improving other aspects of the development workflow. Given the availability and widespread use of \Actions{} for CI/CD, projects considering automating this part of the workflow should consider adopting \Actions{}. Projects that use existing tools should become aware that they may need to migrate to \Actions{} at some point.

\textbf{\Actions{} are still not optimal.} When looking for references to \Actions{} in the projects, the most common type of message we found was requests for help. Developers were soliciting help in configuring a particular GitHub Action or mentioning that automation was not working as intended. Projects should be aware that, as often occurs with novel technologies or features, \Actions{} can introduce unforeseen problems. Projects should be prepared to assist developers in debugging and configuring \Actions{} they adopt. Our results also reveal that the use of \Actions{} sometimes makes debugging more difficult, since developers cannot reproduce locally issues related to \Actions{}.

\textbf{Project activity changes with the introduction of \Actions{}.} The adoption of new technology can bring unanticipated consequences to group behavior~\cite{healy2012unanticipated}. According to \citet{mulder2013impact}, many effects are not directly caused by the new technology itself but by the changes in human behavior that it provokes. For example, with the automation of repetitive tasks, human developers can focus on other tasks, which may help explain some of the changes we observed after the adoption of \Actions{}. Our results suggest that the introduction of \Actions{} causes changes in several activity indicators. In particular, we noted fewer accepted pull requests, with fewer commits and more communication, and more rejected pull requests, with fewer comments and more commits. \Actions{} can also introduce a secondary evaluation step to the pull request. Especially at the beginning of the adoption, the number of commits may increase due to the need to meet all requirements imposed by the \Actions{}. Our results may also imply possible negative consequences. \Actions{} may change the discussion patterns in the project. Utility actions, for example, may lead developers to discuss more. Thus, practitioners, who may already handle a high amount of messages in their repositories, must be aware that introducing some actions may increase the number of messages even more. Additional effort is also necessary to investigate the impact on newcomers, who already face a variety of barriers~\cite{balali2018newcomers,steinmacher2015social} and may suffer from the disturbance in communication.
For newcomers, interacting with \Actions{} can be inconvenient, leading developers to lose motivation or even abandon their contributions. Similar effects have been observed when newcomers interact with other automation tools, such as development bots, which are often perceived as disruptive and noisy~\cite{wessel2021don}. Therefore, designers should envision automation tools as socio-technical rather than purely technical applications, considering human interaction, developers' collaboration, and ethical concerns~\citep{Storey2016}. The literature still lacks design strategies that include end-user perspectives to enhance the interplay between automation tools and developers on social coding platforms. Future work can devise guidelines and best practices about how to build \Actions{} and adopt them in projects to holistically consider the dynamics of the project. Considering different cognitive styles and preferences may also be the subject of future research \cite{santos2023designing}.

\textbf{Distinguishing human and \Actions{} contributions in empirical studies.} To enable large-scale empirical studies on the usage of automation workflows (i.e., bots, \Actions{}) in social coding platforms, it is necessary to determine which projects rely on this automation and which user accounts work as proxies for automation tools. Several bot detection techniques have been proposed to automatically identify bot contributions in software repositories~\cite{golzadeh2020groundtruth,AbdellatifBotHunter2022,BIMAN}. These techniques usually rely on profile information, account activity, and comment patterns in issue and pull request comments. One of the biggest challenges with identifying automated contributions made by bots remains the occurrence of mixed accounts used by both humans and bots. Since \Actions{} can also be implemented to act on behalf of a regular GitHub user account (i.e., a mixed account), the outcome of empirical analyses may be affected if these accounts are not properly identified.

\section{Limitations and Threats to Validity}

This section discusses the limitations and threats to validity and how we have mitigated them.

\textbf{Generalizability:} Since we selected top-starred software projects, our findings might not be generalized to other or all GitHub projects. In particular, our work focused on open-source repositories. Since the usage of Actions might slightly differ for closed-source projects, our findings might also not be generalized to closed-source, private, or industry repositories on GitHub. One way to overcome this threat is by studying less popular projects hosted on GitHub and also projects that are not open-source. Additionally, even though we considered a large number of projects and our results indicate general trends, we recommend running segmented analyses when applying our results to a given project.

\textbf{Reliability of Results:} To ensure consistency and improve the reliability of our qualitative findings, we have calculated the inter-rater agreement. After achieving an `almost perfect' agreement (Cohen's $\kappa = 0.939$~\cite{mchugh2012interrater}), the disagreements between the two researchers who coded the developer's conversations (Discussion threads and conversations on Discord) have been extensively discussed throughout multiple meetings to reach an agreement.

\textbf{Construct Validity:} As stated by Kalliamvakou et al.~\cite{Kalliamvakou2014}, many merged pull requests appear non-merged. Since we consider the number of merged pull requests, our results may be affected by this threat. Our study can be replicated when automated ways of detecting this issue are developed.

\textbf{Internal Validity:} We applied multiple data filtering steps to the statistical models to reduce internal threats. We varied the data filtering criteria to confirm the robustness of our models. For example, we filtered projects that did not receive pull requests in all months and observed similar phenomena. We also carried out a series of placebo tests~\cite{imbens2008regression} using the same model with the adoption artificially set to different dates to confirm the model's robustness. The assumption of exogeneity of the treatment might be a threat. Another internal limitation of our analysis is that a single project on GitHub might have more than one Action in its workflow and, thus, would be considered twice in our models. Previous research has highlighted that many social and technical aspects affect the pull request acceptance 
\cite{tsay2014influence,dey2020effect}. Such aspects might act as potential confounding effects on our models. Following previous work that also considered interventions to pull requests~\cite{cassee2020silent,wessel2020effects,wessel2021quality,kinsman2021}, we added a set of six control variables, including the total number of pull request authors (as a proxy to the community size), the total number of commits (as a proxy to the activity level), time since the first pull request (capturing the pull request usage maturity) that might influence the independent variables to reduce confounding factors. However, in addition to the already identified variables, other factors might influence the results, and further research is necessary to establish causal relations.

\section{Conclusion}

In this paper, we investigate how software developers use \Actions{} to automate their workflows, how they discuss these \Actions{}, and the effects of the adoption of \Actions{} on pull request dynamics. We collected and analyzed data from 5,000 active GitHub repositories. To understand the impact on practice, we statistically analyzed a sample of 662 open-source projects hosted on GitHub.

Firstly, the findings showed that circa 30\% repositories used \Actions{}. We also found that 973 unique predefined \Actions{} were used within the workflows. Further, we collected and analyzed \Actions{} related discussions and chat excerpts on Discord and found that most of them were related to developers asking for help. These findings indicate that \Actions{} can introduce additional issues related to debugging and contributing. By modeling the data around the introduction of \Actions{}, we noticed different results between merged and non-merged pull requests. For merged pull requests, the number of pull requests and commits decreased while comments increased, and for non-merged pull requests the number of pull requests and commits increased while the number of comments decreasesd.

Practitioners need to make informed decisions about whether to adopt \Actions{} into their projects and how to use them effectively. \Actions{} might allow them to automate repetitive tasks in their projects with their own custom Action. \Actions{} provides hundreds of different \Actions{}, potentially making it difficult for practitioners to decide which Action to use. Our work provides empirical data on which \Actions{} are currently used and how they can impact development processes. Learning from those adopters can provide insights to assist the open-source community in deciding whether to use \Actions{} and how to use them effectively. Future work includes the qualitative investigation of the effects of adopting \Actions{} and the expansion of our analysis for considering the effects of different types of \Actions{} and activity indicators.

\begin{acknowledgements}
This work was partially supported by the NSF grants 1815503, 1900903, 2236198, 2247929, 2303042, and the Australian Research Council's Discovery Early Career Researcher Award (DECRA) funding scheme (DE180100153). We thank Timothy Kinsman for his participation in the initial stage of this study.
\end{acknowledgements}

\bibliographystyle{spbasic}
\bibliography{paper}

\begin{thebibliography}{61}
\providecommand{\natexlab}[1]{#1}
\providecommand{\url}[1]{{#1}}
\providecommand{\urlprefix}{URL }
\expandafter\ifx\csname urlstyle\endcsname\relax
  \providecommand{\doi}[1]{DOI~\discretionary{}{}{}#1}\else
  \providecommand{\doi}{DOI~\discretionary{}{}{}\begingroup
  \urlstyle{rm}\Url}\fi
\providecommand{\eprint}[2][]{\url{#2}}

\bibitem[{Abdellatif et~al.(2022)Abdellatif, Wessel, Steinmacher, Gerosa, and
  Shihab}]{AbdellatifBotHunter2022}
Abdellatif A, Wessel M, Steinmacher I, Gerosa MA, Shihab E (2022) {BotHunter}:
  An approach to detect software bots in {GitHub}. In: International Conference
  on Mining Software Repositories ({MSR}), IEEE Computer Society, pp 6--17,
  \doi{10.1145/3524842.3527959}

\bibitem[{Aniche et~al.(2016{\natexlab{a}})Aniche, Bavota, Treude, Van~Deursen,
  and Gerosa}]{aniche2016validated}
Aniche M, Bavota G, Treude C, Van~Deursen A, Gerosa MA (2016{\natexlab{a}}) A
  validated set of smells in model-view-controller architectures. In: 2016 IEEE
  International Conference on Software Maintenance and Evolution (ICSME), IEEE,
  pp 233--243

\bibitem[{Aniche et~al.(2016{\natexlab{b}})Aniche, Treude, Zaidman,
  Van~Deursen, and Gerosa}]{aniche2016satt}
Aniche M, Treude C, Zaidman A, Van~Deursen A, Gerosa MA (2016{\natexlab{b}})
  {SATT}: Tailoring code metric thresholds for different software
  architectures. In: 2016 IEEE 16th international working conference on source
  code analysis and manipulation (SCAM), IEEE, pp 41--50

\bibitem[{Balali et~al.(2018)Balali, Steinmacher, Annamalai, Sarma, and
  Gerosa}]{balali2018newcomers}
Balali S, Steinmacher I, Annamalai U, Sarma A, Gerosa MA (2018) Newcomers’
  barriers... is that all? an analysis of mentors’ and newcomers’ barriers
  in {OSS} projects. Computer Supported Cooperative Work (CSCW) 27(3):679--714

\bibitem[{Benjamini and Hochberg(1995)}]{benjamini1995controlling}
Benjamini Y, Hochberg Y (1995) Controlling the false discovery rate: a
  practical and powerful approach to multiple testing. Journal of the Royal
  statistical society: series B (Methodological) 57(1):289--300

\bibitem[{Brown and Parnin(2019)}]{Brown2019}
Brown C, Parnin C (2019) Sorry to bother you: Designing bots for effective
  recommendations. In: Proceedings of the 1st International Workshop on Bots in
  Software Engineering, IEEE Press, BotSE '19, p 54–58,
  \doi{10.1109/BotSE.2019.00021}

\bibitem[{Calefato et~al.(2022)Calefato, Lanubile, and
  Quaranta}]{10.1145/3544902.3546636}
Calefato F, Lanubile F, Quaranta L (2022) A preliminary investigation of
  {MLOps} practices in {GitHub}. In: Proceedings of the 16th ACM / IEEE
  International Symposium on Empirical Software Engineering and Measurement,
  Association for Computing Machinery, New York, NY, USA, ESEM '22, p
  283–288, \doi{10.1145/3544902.3546636},
  \urlprefix\url{https://doi.org/10.1145/3544902.3546636}

\bibitem[{Cassee et~al.(2020)Cassee, Vasilescu, and
  Serebrenik}]{cassee2020silent}
Cassee N, Vasilescu B, Serebrenik A (2020) The silent helper: the impact of
  continuous integration on code reviews. In: 27th IEEE International
  Conference on Software Analysis, Evolution and Reengineering, IEEE Computer
  Society

\bibitem[{Chen et~al.(2001)Chen, {Fuchs}, and Chung}]{940726}
Chen SK, {Fuchs} WK, Chung JY (2001) Reversible debugging using program
  instrumentation. IEEE Transactions on Software Engineering 27(8):715--727,
  \doi{10.1109/32.940726}

\bibitem[{Chen et~al.(2021)Chen, Zhang, Chen, Wang, and Wu}]{chen2021let}
Chen T, Zhang Y, Chen S, Wang T, Wu Y (2021) Let's supercharge the workflows:
  An empirical study of {GitHub} actions. In: 2021 IEEE 21st International
  Conference on Software Quality, Reliability and Security Companion (QRS-C),
  IEEE, pp 01--10

\bibitem[{Cook and Campbell(1979)}]{quasiexperimentation}
Cook T, Campbell D (1979) Quasi-Experimentation: Design and Analysis Issues for
  Field Settings. Houghton Mifflin

\bibitem[{Cordeiro et~al.(2021)Cordeiro, Silva, Teixeira, Miranda, and
  d’Amorim}]{cordeiro2021shaker}
Cordeiro M, Silva D, Teixeira L, Miranda B, d’Amorim M (2021) Shaker: a tool
  for detecting more flaky tests faster. In: 2021 36th IEEE/ACM International
  Conference on Automated Software Engineering (ASE), IEEE, pp 1281--1285

\bibitem[{Dabbish et~al.(2012)Dabbish, Stuart, Tsay, and
  Herbsleb}]{Dabbish2012}
Dabbish L, Stuart C, Tsay J, Herbsleb J (2012) Social coding in {GitHub}:
  Transparency and collaboration in an open software repository. In:
  Proceedings of the ACM 2012 Conference on Computer Supported Cooperative
  Work, ACM, New York, NY, USA, CSCW '12, pp 1277--1286,
  \doi{10.1145/2145204.2145396}

\bibitem[{Decan et~al.(2022)Decan, Mens, Mazrae, and Golzadeh}]{decan22actions}
Decan A, Mens T, Mazrae PR, Golzadeh M (2022) On the use of {GitHub} actions in
  software development repositories. In: 2022 IEEE International Conference on
  Software Maintenance and Evolution (ICSME), pp 235--245,
  \doi{10.1109/ICSME55016.2022.00029}

\bibitem[{Dey and Mockus(2020)}]{dey2020effect}
Dey T, Mockus A (2020) Effect of technical and social factors on pull request
  quality for the npm ecosystem. In: Proceedings of the 14th ACM/IEEE
  International Symposium on Empirical Software Engineering and Measurement
  (ESEM), pp 1--11

\bibitem[{Dey et~al.(2020)Dey, Mousavi, Ponce, Fry, Vasilescu, Filippova, and
  Mockus}]{BIMAN}
Dey T, Mousavi S, Ponce E, Fry T, Vasilescu B, Filippova A, Mockus A (2020)
  Detecting and characterizing bots that commit code. In: 17th International
  Conference on Mining Software Repositories ({MSR}), ACM, pp 209--219,
  \doi{10.1145/3379597.3387478}

\bibitem[{Duvall et~al.(2007)Duvall, Matyas, Duvall, and
  Glover}]{duvall2007continuous}
Duvall P, Matyas S, Duvall P, Glover A (2007) Continuous Integration: Improving
  Software Quality and Reducing Risk. A Martin Fowler signature book,
  Addison-Wesley

\bibitem[{Erlenhov et~al.(2019)Erlenhov, de~Oliveira~Neto, Scandariato, and
  Leitner}]{Erlenhov2019}
Erlenhov L, de~Oliveira~Neto FG, Scandariato R, Leitner P (2019) Current and
  future bots in software development. In: Proceedings of the 1st International
  Workshop on Bots in Software Engineering, IEEE Press, BotSE '19, p 7–11,
  \doi{10.1109/BotSE.2019.00009}

\bibitem[{Ga{\l}ecki and Burzykowski(2013)}]{galecki2013linear}
Ga{\l}ecki A, Burzykowski T (2013) Linear mixed-effects models using R: A
  step-by-step approach. Springer Science \& Business Media

\bibitem[{Golzadeh et~al.(2020)Golzadeh, Decan, Legay, and
  Mens}]{golzadeh2020groundtruth}
Golzadeh M, Decan A, Legay D, Mens T (2020) A ground-truth dataset and
  classification model for detecting bots in {GitHub} issue and {PR} comments.
  \eprint{2010.03303}

\bibitem[{Golzadeh et~al.(2022)Golzadeh, Decan, and Mens}]{golzadehrise}
Golzadeh M, Decan A, Mens T (2022) On the rise and fall of {CI} services in
  {GitHub}. In: 2022 IEEE 29th International Conference on Software Analysis,
  Evolution and Reengineering (SANER)

\bibitem[{Gousios et~al.(2014)Gousios, Pinzger, and van
  Deursen}]{gousios2014exploratory}
Gousios G, Pinzger M, van Deursen A (2014) An exploratory study of the
  pull-based software development model. In: Proceedings of the 36th
  International Conference on Software Engineering, ACM, pp 345--355

\bibitem[{Gousios et~al.(2016)Gousios, Storey, and Bacchelli}]{Gousios2016}
Gousios G, Storey MA, Bacchelli A (2016) Work practices and challenges in
  pull-based development: The contributor's perspective. In: Proceedings of the
  38th International Conference on Software Engineering, ACM, New York, NY,
  USA, ICSE '16, pp 285--296, \doi{10.1145/2884781.2884826}

\bibitem[{Hata et~al.(2022)Hata, Novielli, Baltes, Kula, and
  Treude}]{hata2022github}
Hata H, Novielli N, Baltes S, Kula RG, Treude C (2022) {GitHub} discussions: An
  exploratory study of early adoption. Empirical Software Engineering
  27(1):1--32

\bibitem[{Healy(2012)}]{healy2012unanticipated}
Healy T (2012) The unanticipated consequences of technology. Nanotechnology:
  ethical and social Implications pp 155--173

\bibitem[{Hilton(2016)}]{Hilton2016}
Hilton M (2016) Understanding and improving continuous integration. In:
  Proceedings of the 2016 24th ACM SIGSOFT International Symposium on
  Foundations of Software Engineering, Association for Computing Machinery, New
  York, NY, USA, FSE 2016, p 1066–1067, \doi{10.1145/2950290.2983952}

\bibitem[{Hu and Gehringer(2019)}]{Hu2019}
Hu Z, Gehringer E (2019) Use bots to improve {GitHub} pull-request feedback.
  In: Proceedings of the 50th ACM Technical Symposium on Computer Science
  Education, Association for Computing Machinery, New York, NY, USA, SIGCSE
  '19, p 1262–1263, \doi{10.1145/3287324.3293787}

\bibitem[{Imbens and Lemieux(2008)}]{imbens2008regression}
Imbens GW, Lemieux T (2008) Regression discontinuity designs: A guide to
  practice. Journal of econometrics 142(2):615--635

\bibitem[{Kalliamvakou et~al.(2014)Kalliamvakou, Gousios, Blincoe, Singer,
  German, and Damian}]{Kalliamvakou2014}
Kalliamvakou E, Gousios G, Blincoe K, Singer L, German DM, Damian D (2014) The
  promises and perils of mining {GitHub}. In: Proceedings of the 11th Working
  Conference on Mining Software Repositories, ACM, New York, NY, USA, MSR 2014,
  pp 92--101, \doi{10.1145/2597073.2597074}

\bibitem[{Kavaler et~al.(2019)Kavaler, Trockman, Vasilescu, and
  Filkov}]{kavaler2019tool}
Kavaler D, Trockman A, Vasilescu B, Filkov V (2019) Tool choice matters:
  {JavaScript} quality assurance tools and usage outcomes in {GitHub} projects.
  In: Proceedings of the 41st International Conference on Software Engineering,
  IEEE Press, pp 476--487

\bibitem[{Kinsman et~al.(2021)Kinsman, Wessel, Gerosa, and
  Treude}]{kinsman2021}
Kinsman T, Wessel M, Gerosa M, Treude C (2021) How do software developers use
  {GitHub} actions to automate their workflows? In: Mining Software
  Repositories Conference (MSR), IEEE

\bibitem[{Kuznetsova et~al.(2017)Kuznetsova, Brockhoff, and
  Christensen}]{kuznetsova2017lmertest}
Kuznetsova A, Brockhoff PB, Christensen RHB (2017) lmertest package: tests in
  linear mixed effects models. Journal of Statistical Software 82(13)

\bibitem[{Leite et~al.(2013)Leite, Ansaldi~Oliva, Nogueira, Gerosa, Kon, and
  Milojicic}]{leite2013systematic}
Leite LA, Ansaldi~Oliva G, Nogueira GM, Gerosa MA, Kon F, Milojicic DS (2013) A
  systematic literature review of service choreography adaptation. Service
  Oriented Computing and Applications 7:199--216

\bibitem[{Lin et~al.(2016)Lin, Zagalsky, Storey, and Serebrenik}]{Lin2016}
Lin B, Zagalsky A, Storey M, Serebrenik A (2016) Why developers are slacking
  off: Understanding how software teams use {Slack}. In: Proceedings of the
  19th ACM Conference on Computer Supported Cooperative Work and Social
  Computing Companion, ACM, New York, NY, USA, CSCW '16 Companion, pp 333--336,
  \doi{10.1145/2818052.2869117}

\bibitem[{McHugh(2012)}]{mchugh2012interrater}
McHugh ML (2012) Interrater reliability: the kappa statistic. Biochemia medica
  22(3):276--282

\bibitem[{Mirhosseini and Parnin(2017)}]{Mirhosseini2017}
Mirhosseini S, Parnin C (2017) Can automated pull requests encourage software
  developers to upgrade out-of-date dependencies? In: Proceedings of the 32nd
  IEEE/ACM International Conference on Automated Software Engineering, IEEE
  Press, ASE 2017, p 84–94

\bibitem[{Monperrus(2019)}]{Monperrus2019}
Monperrus M (2019) Explainable software bot contributions: Case study of
  automated bug fixes. In: Proceedings of the 1st International Workshop on
  Bots in Software Engineering, IEEE Press, Piscataway, NJ, USA, BotSE '19, pp
  12--15, \doi{10.1109/BotSE.2019.00010}

\bibitem[{Mulder(2013)}]{mulder2013impact}
Mulder K (2013) Impact of new technologies: how to assess the intended and
  unintended effects of new technologies. Handb Sustain Eng(2013)

\bibitem[{Nakagawa and Schielzeth(2013)}]{nakagawa2013general}
Nakagawa S, Schielzeth H (2013) A general and simple method for obtaining {R2}
  from generalized linear mixed-effects models. Methods in ecology and
  evolution 4(2):133--142

\bibitem[{Ouyang et~al.(2006)Ouyang, Dumas, Ter~Hofstede, and Van~der
  Aalst}]{ouyang2006bpmn}
Ouyang C, Dumas M, Ter~Hofstede AH, Van~der Aalst WM (2006) From {BPMN} process
  models to {BPEL} web services. In: 2006 IEEE International Conference on Web
  Services (ICWS'06), IEEE, pp 285--292

\bibitem[{Pinto et~al.(2017)Pinto, Figueira~Filho, Steinmacher, and
  Gerosa}]{pinto2017training}
Pinto GHL, Figueira~Filho F, Steinmacher I, Gerosa MA (2017) Training software
  engineers using open-source software: the professors' perspective. In: 2017
  IEEE 30th Conference on Software Engineering Education and Training
  (CSEE\&T), IEEE, pp 117--121

\bibitem[{Santos et~al.(2023)Santos, Pimentel, Wiese, Steinmacher, Sarma, and
  Gerosa}]{santos2023designing}
Santos I, Pimentel JF, Wiese I, Steinmacher I, Sarma A, Gerosa MA (2023)
  Designing for cognitive diversity: Improving the github experience for
  newcomers. In: Proceedings of the 2023 ACM/IEEE 45th International Conference
  on Software Engineering: Software Engineering in Society

\bibitem[{dos Santos and Gerosa(2018)}]{dos2018impacts}
dos Santos RM, Gerosa MA (2018) Impacts of coding practices on readability. In:
  Proceedings of the 26th Conference on Program Comprehension, pp 277--285

\bibitem[{Saroar and Nayebi(2023)}]{saroar23}
Saroar SG, Nayebi M (2023) Developers’ perception of {GitHub} actions: A
  survey analysis. In: 2023 International Conference on Evaluation and
  Assessment in Software Engineering (EASE)

\bibitem[{Sheather(2009)}]{sheather2009modern}
Sheather S (2009) A modern approach to regression with R. Springer Science \&
  Business Media

\bibitem[{Steinmacher et~al.(2015)Steinmacher, Conte, Gerosa, and
  Redmiles}]{steinmacher2015social}
Steinmacher I, Conte T, Gerosa MA, Redmiles D (2015) Social barriers faced by
  newcomers placing their first contribution in open source software projects.
  In: Proceedings of the 18th ACM conference on Computer supported cooperative
  work \& social computing, pp 1379--1392

\bibitem[{Storey and Zagalsky(2016)}]{Storey2016}
Storey MA, Zagalsky A (2016) Disrupting developer productivity one bot at a
  time. In: Proceedings of the 2016 24th ACM SIGSOFT International Symposium on
  Foundations of Software Engineering, ACM, New York, NY, USA, FSE 2016, pp
  928--931, \doi{10.1145/2950290.2983989}

\bibitem[{Subash et~al.(2022)Subash, Kumar, Vadlamani, Chatterjee, and
  Baysal}]{subash2022disco}
Subash KM, Kumar LP, Vadlamani SL, Chatterjee P, Baysal O (2022) Disco: A
  dataset of {Discord} chat conversations for software engineering research.
  In: Proceedings of the 19th International Conference on Mining Software
  Repositories, pp 227--231

\bibitem[{Thistlethwaite and Campbell(1960)}]{thistlethwaite1960regression}
Thistlethwaite DL, Campbell DT (1960) Regression-discontinuity analysis: An
  alternative to the ex post facto experiment. Journal of Educational
  psychology 51(6):309

\bibitem[{van Tonder and Goues(2019)}]{Tonder2019}
van Tonder R, Goues CL (2019) Towards s/engineer/bot: Principles for program
  repair bots. In: Proceedings of the 1st International Workshop on Bots in
  Software Engineering, IEEE Press, BotSE '19, p 43–47,
  \doi{10.1109/BotSE.2019.00019}

\bibitem[{Tsay et~al.(2014)Tsay, Dabbish, and Herbsleb}]{tsay2014influence}
Tsay J, Dabbish L, Herbsleb J (2014) Influence of social and technical factors
  for evaluating contribution in {GitHub}. In: Proceedings of the 36th
  international conference on Software engineering, pp 356--366

\bibitem[{Valenzuela-Toledo and Bergel(2022)}]{valenzuelaevolution}
Valenzuela-Toledo P, Bergel A (2022) Evolution of {GitHub} action workflows.
  In: 2022 IEEE 29th International Conference on Software Analysis, Evolution
  and Reengineering -- Early Research Achievements (SANER-ERA)

\bibitem[{Vasilescu et~al.(2015)Vasilescu, Yu, Wang, Devanbu, and
  Filkov}]{Vasilescu2015}
Vasilescu B, Yu Y, Wang H, Devanbu P, Filkov V (2015) Quality and productivity
  outcomes relating to continuous integration in {GitHub}. In: Proceedings of
  the 2015 10th Joint Meeting on Foundations of Software Engineering, ACM, New
  York, NY, USA, ESEC/FSE 2015, pp 805--816, \doi{10.1145/2786805.2786850}

\bibitem[{Wessel and Steinmacher(2020)}]{wessel2020inconvenient}
Wessel M, Steinmacher I (2020) The inconvenient side of software bots on pull
  requests. In: Proceedings of the IEEE/ACM 42nd International Conference on
  Software Engineering Workshops, Association for Computing Machinery, New
  York, NY, USA, ICSEW'20, p 51–55, \doi{10.1145/3387940.3391504}

\bibitem[{Wessel et~al.(2018)Wessel, de~Souza, Steinmacher, Wiese, Polato,
  Chaves, and Gerosa}]{Wessel2018}
Wessel M, de~Souza BM, Steinmacher I, Wiese IS, Polato I, Chaves AP, Gerosa MA
  (2018) The power of bots: Characterizing and understanding bots in {OSS}
  projects. Proc ACM Hum-Comput Interact 2(CSCW):182:1--182:19,
  \doi{10.1145/3274451}

\bibitem[{{Wessel} et~al.(2020){Wessel}, {Serebrenik}, {Wiese}, {Steinmacher},
  and {Gerosa}}]{wessel2020effects}
{Wessel} M, {Serebrenik} A, {Wiese} I, {Steinmacher} I, {Gerosa} MA (2020)
  Effects of adopting code review bots on pull requests to {OSS} projects. In:
  2020 IEEE International Conference on Software Maintenance and Evolution
  (ICSME), pp 1--11, \doi{10.1109/ICSME46990.2020.00011}

\bibitem[{Wessel et~al.(2021)Wessel, Wiese, Steinmacher, and
  Gerosa}]{wessel2021don}
Wessel M, Wiese I, Steinmacher I, Gerosa MA (2021) Don't disturb me: Challenges
  of interacting with software bots on open source software projects.
  Proceedings of the ACM on Human-Computer Interaction 5(CSCW2):1--21

\bibitem[{Wessel et~al.(2022{\natexlab{a}})Wessel, Abdellatif, Wiese, Conte,
  Shihab, Gerosa, and Steinmacher}]{wessel2022icse}
Wessel M, Abdellatif A, Wiese I, Conte T, Shihab E, Gerosa MA, Steinmacher I
  (2022{\natexlab{a}}) Bots for pull requests: The good, the bad, and the
  promising. In: Proceedings of the 44th International Conference on Software
  Engineering, Association for Computing Machinery, New York, NY, USA, ICSE
  '22, p 274–286, \doi{10.1145/3510003.3512765},
  \urlprefix\url{https://doi.org/10.1145/3510003.3512765}

\bibitem[{Wessel et~al.(2022{\natexlab{b}})Wessel, Serebrenik, Wiese,
  Steinmacher, and Gerosa}]{wessel2021quality}
Wessel M, Serebrenik A, Wiese I, Steinmacher I, Gerosa MA (2022{\natexlab{b}})
  Quality gatekeepers: Investigating the effects of code review bots on pull
  request activities. Empirical Software Engineering 27(108),
  \doi{10.1007/s10664-022-10130-9}

\bibitem[{Wyrich and Bogner(2019)}]{Wyrich2019}
Wyrich M, Bogner J (2019) Towards an autonomous bot for automatic source code
  refactoring. In: Proceedings of the 1st International Workshop on Bots in
  Software Engineering, IEEE Press, Piscataway, NJ, USA, BotSE '19, pp 24--28,
  \doi{10.1109/BotSE.2019.00015}

\bibitem[{Zhao et~al.(2017)Zhao, Serebrenik, Zhou, Filkov, and
  Vasilescu}]{zhao2017impact}
Zhao Y, Serebrenik A, Zhou Y, Filkov V, Vasilescu B (2017) The impact of
  continuous integration on other software development practices: a large-scale
  empirical study. In: Proceedings of the 32nd IEEE/ACM International
  Conference on Automated Software Engineering, IEEE Press, pp 60--71

\end{thebibliography}

\end{document}